\documentstyle[11pt,aaspp4]{article}
\def\ang{\AA}
\def\etal{{\it et~al.}}
\def\eg{{\it e.g.\/}}

\begin{document}

\title{Redshifts and Optical Properties for a Statistically Complete Sample of Poor Galaxy Clusters}
\author{Michael J. Ledlow, Chris Loken, Jack O. Burns} 
\affil{New Mexico State University, Dept. of Astronomy, Box 30001/Dept. 4500, \\ 
Las Cruces, NM 88003-0001} 
\vskip 0.3in
\author{John M. Hill}
\affil{Steward Observatory, University of Arizona, Tuscon, AZ 85721} 
\vskip 0.3in 
\centerline{and}
\author{Richard A. White}
\affil{NASA/Goddard Space Flight Ctr., Space Data and Computing Division, 
Code 932, Greenbelt, MD 20771} 

\begin{abstract}
     
     From the poor cluster catalog of White \etal\ (1996), 
we define a sample of 71 optically-
selected poor galaxy clusters.  The surface-density enhancement we require 
for our clusters falls between that of the loose associations of Turner and Gott
(1976) and the Hickson compact groups (Hickson, 1982). 
We review the selection biases and 
determine the statistical completeness of the sample.  For this sample, 
we report new 
velocity measurements made with the ARC 3.5-m Dual-Imaging spectrograph 
and the 2.3-m
Steward Observatory MX fiber spectrograph.  Combining our own measurements
with those from the literature, we examine the velocity distributions,
velocity dispersions, and 1-d velocity substructure for our poor cluster
sample, and compare our results to other poor cluster samples.  
We find that approximately half of the sample may have significant 
1-d velocity substructure.  The optical morphology, large-scale environment, 
and velocity field of many of these clusters are indicative of young, dynamically
evolving systems. 
In future papers, we will use this sample to derive the poor cluster X-ray 
luminosity function and gas mass function, and will examine the optical/X-ray properties
of the clusters in more detail. 
\end{abstract}


\clearpage
\section{Introduction}

     Galaxy groups or poor clusters are interesting for many reasons.
Poor clusters are much more numerous than rich clusters like those
in the Abell (1958) catalog.  One estimate suggests that over 50\%
of all galaxies reside in a poor cluster environment (Dressler, 1984; 
Soneira \& Peebles, 1978).  
Because of their high space density, 
poor clusters are very good tracers of large-scale structure, and hence,
their properties may provide important constraints to cosmological 
models of large-scale formation.  

     The definition of a poor cluster is rather vague.  In most 
respects, {\it poor} simply means that the number of galaxies falls below 
some limiting (for instance Abell's richness class 0) criteria for a 
{\it rich} cluster.  In practice, there are many existing catalogs
of poor clusters which each differ greatly in their selection criteria, 
such as the degree of isolation, compactness, and galaxy counts.  
Some of the earlier poor cluster catalogs (Shakhbazyan, 1973; Rose, 1977; 
Hickson, 1982) concentrated on 
very compact systems, with surface-density 
enhancements as high as 100-1000 times that of the mean field density
(\eg\ Hickson Compact Groups (HCG)). 
Many of these compact groups have surface densities equal to or greater 
than the centers of many rich clusters.  Conversely, Turner and Gott (1976)
defined a catalog of poor clusters, using a friends-of-friends approach, 
which located many {\it loose} systems with only a 4.6 surface-density 
enhancement over the field.  Other selection criteria, such as the presence
of a dominant cD or D-type first-ranked galaxy (Albert, White, \& Morgan, 1977; Morgan,
Kayser, \& White, 1975) have resulted in other poor cluster samples (the AWM and
MKW clusters).  
Thus, {\it poor clusters} can span the entire
range from isolated pairs of galaxies to just below the threshold of 
rich galaxy clusters.  It is for this reason that poor clusters are so 
valuable for studies of galaxy formation, dynamics, and evolution, as well 
as hierarchical large-scale structure formation and cosmology.  

     The optical properties of various poor cluster samples have been 
studied recently by Beers \etal\ (1995) (MKW and AWM clusters), Diaferio, Geller, \& Ramella 
(1994) (HCG), Dell'Antonio, Geller, \& Fabricant  
(1994) (MKW, AWM, and White \etal\ 1996 clusters),  
and Ramella \etal\ (1994) (HCG's and CfA loose groups).  
The existence of very compact groups is particularly troublesome because 
of their short crossing times ($t_{cr} \leq 0.02~{H_0}^{-1}$) 
(Diaferio, Geller, \& Ramella 1994).
Dynamical friction should cause the galaxies to merge into a single system 
on a time-scale of order 5-10 crossing times (the merging instability, \eg\ Barnes 1985).
Thus, it is 
surprising that we see any very compact groups at all. If the merging scenario 
does occur in compact groups, the AWM and MKW clusters may be good candidates
for the end-result, culminating in the formation of 
a very extended central-dominant galaxy.  Beers \etal\ used this idea to 
look for D/cD galaxies in MKW and AWM clusters with large peculiar velocities
relative to the cluster mean in order to constrain dynamical ages and 
merging time-scales.  They found 4/29 clusters with possibly significant 
D/cD velocity offsets which may be in the process of forming the dominant
galaxy via the merger of member galaxies.  
The lifetime
problem may be resolved, however, if compact systems are continually 
forming within 
collapsing, rich, loose clusters (Ramella \etal\ 1994).  Beers \etal\, Ramella
\etal\, and Diaferio \etal\ all show evidence that compact groups are often
imbedded within, and often gravitationally bound to much larger galaxy 
associations.  There is substantial evidence to suggest that loose/compact
groups may be related by similar formation mechanisms or via an evolutionary 
sequence from loose to compact systems.  Thus, dynamical studies of 
large samples of poor groups are very important to better our understanding
of how clusters form and evolve.  

     What is the relationship between poor and rich clusters?  X-ray 
studies have shown that most poor groups are in fact real galaxy associations
with a detectable hot intragalactic medium (IGM) with T $\sim$ 1 keV ({\it e.g.\/} 
Doe \etal\ 1995).  
Relationships
between the X-ray luminosity, temperature and optical measures,
such as the velocity dispersion, suggest that poor clusters are simply
lower mass extensions of richer clusters (Price \etal\ 1991).  However,
there appears to be a turnover in the $L_x$ vs. $\sigma_v$ relationship
as the member galaxies contribute a larger and larger fraction of the
total X-ray emission (Dell'Antonio, Geller, \& Fabricant 1994). 
This relationship 
is also complicated because the velocity dispersion is not necessarily
a good estimator of the mass of the system if the cluster is not relaxed
(Diaferio, Geller, \& Ramella 1995).  Additionally, there is some evidence to 
suggest that the distribution of dark matter, the baryon fraction, and 
mass-to-light ratios may be different from rich clusters (Dell'Antonio, Geller, \& Fabricant 
 1995; David, Jones, \& Forman 1995).  These conclusions are fairly controversial,
however, because data of sufficient quality are only available for a few clusters.
In order to address these questions, one needs large samples.  It is with
these goals in mind that we have defined a new sample of poor clusters for 
a detailed study of their optical, x-ray, and radio properties, and their
connection to existing statistical samples of rich clusters.  

     We describe the selection of the poor cluster sample in section 2.
In section 3, we present the observations and the new velocity measurements
for this sample.  In section 4, we examine the completeness of the 
sample, and select a volume-limited subsample for further study.  In section 5, we  
discuss the velocity distributions, dispersions, and the results of statistical
1-D normality tests.  We discuss the properties of several clusters in 
detail in the appendix. 

     Throughout this paper we assume $\rm H_0 = 75~km~sec^{-1}~Mpc^{-1}$ and
$q_0 = 0.5$.  

\section{Sample Selection} 

     The galaxy clusters were selected from the electronic version of the
Zwicky catalog of galaxies and clusters of galaxies (CGCG) (Zwicky \etal\
1961-1968).  White \etal\ (1996) used a friends-of-friends technique (Turner and Gott
 1976)
to identify galaxy groups based on surface-density enhancements relative
to the north galactic polar cap and included all 
galaxies to the magnitude limit of the survey (15.7).  The catalog was 
searched in $10^{\circ}$ galactic latitude strips with a $1^{\circ}$ overlap
with successive scans in order to avoid missing groups near the edge of 
the strips. 
The base-level 
surface density enhancement of the entire catalog is 21.54 (or $100^{2/3}$, 
equal to a volume-density enhancement of 100)
with a minimum of 3 Zwicky galaxies per group.  This search produced $>600$ potential
groups, including all MKW, 
AWM, WP (White 1978) groups as well as
some nearby Abell clusters (Abell 1958).  Radio observations of a subsample
of this catalog were reported in Burns \etal\ (1987). 
An X-ray/optical study of groups from the Burns \etal\ sample with {\it 
Einstein} 
observations was reported in Price \etal\ (1991).  

    In this paper, and those to follow in this series, we have constructed
a statistically complete subsample of the original White \etal\ catalog
corresponding to those groups with a 46.4 surface-density enhancement 
(a $10^{1/3}$ 
higher volume density over the base-level of the catalog), which included
4 or more Zwicky galaxies, and with galactic latitude $|b|\geq 30^{\circ}$. 
After eliminating several Abell clusters (A119, A194, A400, A634, A779, A1185,
A1213, A1367, A1656, A1983, A2022, the Hercules and Pisces superclusters, A2052, A2063, 
A2199, A2634, and A2666) 
and 4 redundant
groups (those found in more than one search strip) the sample contains 
71 poor clusters. Statistically, approximately 39\% of the poor clusters 
have only the minimum 
number of 4 galaxies down to the magnitude limit of the search.  
The distribution of galaxy 
number has a tail which extends to a maximum of 10, with a mean of 6 galaxies.
The entire poor cluster sample is listed in Table 1.  

\section{Observations} 

     A thorough search of the literature and the NED extragalactic
database revealed that 
only about half of the sample had $>2$ measured velocities
per cluster.  We require a minimum of 2 concordant velocities in order
to reliably determine the distance to the cluster. 
We therefore began a program to measure recessional velocities 
for the 
remainder of these clusters in order to use this sample to study the 
statistical properties
of nearby poor clusters. 

     Spectroscopic observations of the cluster galaxies were obtained using the
Astrophysical Research Consortium's (ARC) 3.5-meter telescope at 
Apache Point, NM.  The instrument
used was the Dual-Imaging Spectrograph (DIS) mounted
at the Nasmyth focus. 
The DIS allows both imaging and spectroscopy modes
by rotating either mirrors or gratings, mounted in a turret wheel, into the
light path.  Two grating sets are available; a low-resolution set 
(150/300 lines mm$^{-1}$) with  6-7\/\AA\ pixel$^{-1}$ spectral resolution, 
and a high-resolution
set (831/1200 lines mm$^{-1}$) at 1-1.7\/\AA\ pixel$^{-1}$.  All observations reported here were
made in low-resolution mode. 

     The incoming light-beam is split by a dichroic filter with a wavelength 
cutoff
near 5700$\ang$.  The blue and red beams are each directed along separate
light-paths to two CCD detectors.  The blue-chip is a TEK 512x512 CCD, 
and the red-chip a TI 800x800 array.  Plate-scales are $\rm 1.086~\arcsec/pixel$
and $\rm 0.61~\arcsec/pixel$ on the blue and red chips, respectively.  The 
field-of-view on both chips is approximately $6\arcmin\times4\arcmin$. A 
filter wheel is mounted behind the dichroic with a Gunn {\it g,r} 
set for imaging mode. 
The spectra are also obtained separately from the 
blue and red chips.  In low-resolution mode, the spectral coverage 
extends from $3800-10000\ang$, divided into two parts blueward and 
redward of $5700\ang$.  
A separate filter wheel is located at the focal-plane
at the entrance to the spectrograph to hold the slit for spectroscopy.

     We report observations made on the nights of 27-28 November 1994, 
24-25 January, 9 March, and 22 May 1995.
On the earlier observing runs, we used a single $2\arcsec$ slit, and
rotated the detector to align two or more galaxies along the slit. In later
runs, we used custom-designed slit masks in order to observe 4-12 
galaxies simultaneously.  Candidate galaxies were identified from images
extracted from the CD-ROM Digitial POSS I Survey.  Positions were measured
with $<1\arcsec$ accuracy, 
and pointing centers were determined by optimizing the maximum number 
of galaxies which could be fit into the DIS field-of-view.  From these 
coordinate files, an aperture slit-mask was designed matching the 
imaging scale of the DIS spectrograph using a
custom-written IDL program.  By default, we used $18\arcsec\times 2\arcsec$
slits oriented East-West.  In some cases, slits were 
located off-center from the galaxy position in order to observe galaxies closer
than $18\arcsec$ separation in right-ascension. 
The final versions of the slit-masks
are saved to a Postscript file and printed on Kodak infrared film on a 
$2500~\rm dpi$ typesetter as a negative transparency. 
Tests
have shown the negative transparency film to be very opaque (no scattered light
problems), and the clear slits to have approximately equal transmission 
as compared to 
the default Quartz-slits used in the DIS.  These slit-masks are cut out
and placed in the filter/slit-wheel at the entrance to the spectrograph.

    The observing procedure consisted of 1) imaging the poor cluster field
through the Gunn {\it g,r} filters in imaging mode,  2) imaging the
slit-mask to locate the positions of the slits on the detector, 
3) using an on-line Fortran program to calculate the necessary right-ascension
and declination offsets and rotation angle in order to align the candidate 
galaxies in the appropriate slits, 4) offsetting the telescope and rotating 
the instrument as determined in (3), and 5) moving the gratings into the 
light-path and beginning the exposure.  Typical exposure times were 10 minutes.
Wavelength calibration was obtained by taking spectra of Helium-Neon
calibration lamps following each object observation. 

     We used standard spectroscopic reduction procedures in IRAF to 
analyze the spectra.  The bias-level was determined from a 30 column overscan
region, and a zero-frame was subtracted to remove large-scale bias fluctuations.
For the single-slit observations, a spectral flatfield was used to correct
for the illumination pattern of the slit and cosmetic defects on the CCD
chips.  For the multislit data, spectral flats were complicated by 
overlapping dispersion patterns from adjacent slits (using 
the continuum lamp), so this correction was not used.  For single-slit
observations, the wavelength solution was determined for both axes of
the 2-d image.  For the multislit data, the wavelength solution was applied
to the background-subtracted and extracted 1-d spectra.  This was necessary
as each spectrum had a unique wavelength range dependent on the 
north-south position of the slit on the chip.  
Heliocentric radial velocities were measured
using the cross-correlation program FXCOR.  We observed several 
nearby galaxies and G and K-type radial velocity standard stars 
as template spectra for the cross-correlations.  For the R-spectra,
the atmospheric A-band was removed by restricting the wavelength
range for the cross-correlation for both the object and template.

     In Figure 1, we show a sample ARC 3.5-m DIS spectrum for IC 4508 
in the poor 
group N67-310 at a redshift of $z=0.045$.  IC 4508 is one of 7 
galaxies observed for this cluster using the slit masks discussed above. 
In Figure 1(a) we show the blue portion of the spectrum from $3800-5700~\ang$.
The red portion of the spectrum in Figure 1(b) extends from $\sim 5700-10500~\ang$. 
The shape of the spectrum redward of $5250\ang$ on the blue spectra, 
and blueward of $6100\ang$ on the red spectra is due to the wavelength
cutoff of the dichroic. 

     In addition to the ARC 3.5-m observations, we also report multifiber 
spectroscopic observations of 3 poor clusters from this statistical 
sample and 3 other poor clusters from the larger sample of 
Burns \etal\ (1987) with ROSAT PSPC observations reported in Doe \etal\ (1995)
plus one other cluster (N03-183) which meets the statistical sample definition,
but is located at low galactic latitude ($23^{\circ}$).  
These data were obtained in January and 
July 1994 using the
University of Arizona's Steward Observatory 2.3-m telescope and the 
MX multifiber spectrograph.  A detailed description of the design and
construction of the MX spectrometer is given by Hill and Lesser (1986). 
The detector was a UV-flooded Loral $1200 \times
800$ pixel CCD with a read-noise of 8 electrons pixel$^{-1}$ rms.  A 
400 line mm$^{-1}$ grating provided wavelength coverage from 3700 to 
6950\AA\ at $8$\AA\ resolution (2.8\/\AA\ pixel$^{-1}$). 

     In Table 2, we list the new ARC heliocentric velocity measurements 
for our poor cluster sample.  We have identified the individual galaxies 
from positions measured from the Digital-Sky Survey and cross-correlated
with NED and the Zwicky catalog.  Some identifications from these 
catalogs corresponded to 
galaxy pairs with a single name.  In these cases, we have used the 
catalogued designation and appended a number to distinguish the objects. 
Hubble types for the galaxies were 
determined from examination of the Gunn {\it g,r} images from DIS. 
Table 3 lists heliocentric velocities measured 
using MX for the clusters S49-128, S34-111, S49-140, S49-132, N56-394, MKW2,  
and N03-183. 

\section{Sample Properties} 

     In order to use this sample of galaxy clusters for statistical studies
and, in particular, for determination of the X-ray luminosity function 
(Burns \etal\ 1996), it is necessary to examine the 
completeness in terms of a volume-limited sample.  In Figure 2, we plot the 
volume density of clusters as a function of redshift.  
We see that for $z>0.03$, the volume density begins to fall off rapidly.
This is a well-known consequence of the magnitude limit of the Zwicky catalog.  
As one goes to higher redshift, the number of group galaxies with 
magnitudes brighter than 15.7 decreases because of the shape of the 
optical luminosity function.  Thus at higher redshift, we are only 
able to pick out clusters with higher richness values and more bright
galaxies.  These clusters may therefore not be representative of the 
sample below $z=0.03$.  Below redshifts of $z=0.01$, there are only 
4 clusters in the catalog, and these are most likely members of the 
Local Group. We eliminated these clusters from the statistical sample 
because the measured velocities may be strongly affected by peculiar
motions.  Within the Poisson counting errors of our sample (49 clusters 
with $0.01 \leq z \leq 0.03$), there is no statistical difference 
between the observed volume-density and a flat distribution with a mean
of $ 2.9\pm 0.3 \times 10^{-5}~Mpc^{-3}$.  
We therefore judge our sample volume-limited within observational uncertainties. 
In Table 1, we have marked those
clusters which fall outside our volume-limited sample with an asterisk.
Our statistical sample includes 49 poor clusters.  
We calculate that our 
survey covers 10973 square degrees (3.34 steradians) or 26.6\% of the sky.  

     The base-level catalog from White \etal\ includes clusters selected
from all galactic latitude strips, limited only by the $-3^\circ$
declination cutoff of the Zwicky catalog.  In the selection of this 
sample, we have limited the galactic latitude range to $|b| \geq 30^{\circ}$.
We investigated whether some incompleteness in the sample could be attributed 
to galactic absorption at the lower latitudes.  Within the error bars, the 
surface density of clusters is constant as a function of galactic latitude.  

     What is the relationship between our poor clusters and Zwicky clusters?
Zwicky did not use a strict definition (as did Abell) for defining clusters.
In fact, his catalog contains clusters with a wide range in properties.
His open clusters are very large structures ($\geq 1^{\circ}$) which may 
contain many smaller subclumps of galaxy groups. 
Surface-density enhancements were also cataloged as either medium compact
or compact.  We performed a cross-correlation between our 67 
(excluding the 4 clusters with $z<0.01$) poor clusters
and the nearby Zwicky clusters.  We find that 53/67 are found within 
catalogued clusters.  Another 6 are not within the Zwicky cluster boundary, 
but are near a cluster, or in a group of clusters.  Eight do not appear
to be associated with any listed Zwicky cluster.  Of the 53 which were
found in Zwicky clusters, 26 are classified as open, 26 as medium compact,
and 1 as compact.  Four of the poor clusters are located within the 
Coma supercluster, and one within the Hercules supercluster.  Also, 
some of Zwicky's clusters contain more than one of our poor clusters 
({\it e.g.\/} N56-359, N56-371, N56-392 all in Zw1141.7--0158, N34-169 and 
N45-388 in 
Zw0909.7+1814).  
The location of our poor clusters within Zwicky's contours is variable,
ranging from centrally located to the edge of the galaxy distribution
(see Doe \etal\ 1995).  

     We compared the 26:26:1 compactness ratio for our sample to the nearby
Zwicky clusters in general.  From Zwicky's catalog, 504 clusters are 
classified as nearby from apparent magnitudes or velocity information.
The ratio of Open:Medium-Compact:Compact is 274:205:25.  From a $2\times2$
contingency table test there is no statistically significant difference
in the ratios between our sample and the nearby Zwicky catalog. Thus our 
sample appears to be representative of Zwicky clusters in general. 
It is interesting that nearly half of our sample is located within 
Zwicky's open clusters. 
The statistics described above also suggest that
a large fraction of our clusters are correlated with or are embedded within
larger-scale 
structures.  We will examine this question with regard to the velocity
field and velocity dispersions in the next section.

\section{Velocity Distributions and Dispersions} 

     Of the 71 clusters listed in Table 1, 40 have $\geq 5$ velocity 
measurements within a 0.5 Mpc projected radius from the optical center of the
cluster. 
We report velocity 
dispersions for all clusters with $\geq 5$ measurements in column 6.
We have calculated the Biweight estimators of location and scale 
($C_{BI}$ and $S_{BI}$) (Beers, Flynn, \& Gebhardt 1990) 
rather than the average and standard deviation (Danese, DeZotti, \& di Tullio 1980).
These quantities are 
less sensitive to small number statistics and outliers in the velocity 
distribution.  In addition, we list the $1\sigma$ confidence invervals 
determined from the Jackknife
method (Mosteller \& Tukey 1977).  
For those clusters which had at least a 50\% increase in the number
of velocities out to 1 Mpc, we also list the velocity dispersion 
separately for $R \leq 1~Mpc$ in column 8. 

     In Figure 3, we plot a histogram of the velocity dispersions
measured within 0.5 Mpc.  The median dispersion 
is $295\pm 31~\rm km~sec^{-1}$ (using the error on the median). 
Approximately 17\% of the clusters
fall in the tail of the distribution, with $\sigma_v > 500~\rm km~sec^{-1}$. 
Zabludoff, Huchra, \& Geller (1990) reported an 
average $\sigma_v=744~\rm km~sec^{-1}$ for a sample of nearby rich clusters.
Thus, there is 
a small amount of overlap in the long-tail of our distribution with
richer clusters.  Ramella \etal\ (1994) compared the velocity 
dispersions of 29 HCG's to 36 loose systems (RGH) selected from the CfA 
redshift survey. 
They calculated median values of $329\pm 135$ and $403\pm 96$ for
the HCG and RGH clusters, respectively.  For 21 MKW and AWM clusters,
Beers \etal\ (1995) found a median dispersion of $336\pm 40~\rm km~sec^{-1}$.
These values are consistent
with our sample, although it should be noted that both Ramella \etal\ 
and Beers \etal\ included
galaxies out to a larger radius (1.125 Mpc for $H_0=75$).  

     Diaferio \etal\ (1994) computed velocity dispersions for a number 
of simulated compact groups using an N-body Treecode 
(Hernquist, 1987; Barnes \& 
Hut 1986).  They found a median $\sigma_v=257~\rm km~sec^{-1}$ for
their simulated clusters.  While this is lower than 
our value of $295~\rm km~sec^{-1}$, given the large spread in our 
distribution, the results may be consistent.  
Our observed distribution has $\sim 17\%$
more clusters with $S_{BI} \geq 400~km~sec^{-1}$ than predicted by the 
simulations. However, given the errors on the velocity dispersions, 
it is unlikely that this result is statistically significant. 
We performed a KS-test 
and T-test to compare the $S_{BI}$ distribution in Figure 3 to the 
simulated data from Diaferio \etal\ (Figure 8(a) in their paper).  
We can reject the hypothesis that the two distributions are drawn from 
the same parent population at no better than the 20\% level. 

     We have compared the velocity dispersions for $r\leq 0.5~Mpc$, $r \leq 1~Mpc$, 
and $0.5~Mpc~\leq r \leq 1~Mpc$.  Because of the often low number of velocities, the errors on the dispersions are quite large.  We find that a histogram 
of the difference in $S_{BI}$ for different radii is consistent with a 
Gaussian spread of values with FWHM $\sim 200~km~sec^{-1}$.  The best-fit Gaussian is 
slightly skewed towards higher $S_{BI}$ ($\Delta S_{BI} \sim 75~km~sec^{-1}$). 
Within the errorbars, however, the significance of the difference from zero
is inconclusive. 
Ramella \etal\ (1994) claimed that in general, the velocity dispersion increases 
when one includes the larger association around HCG's.  Such analysis has the 
potential of providing important dynamical information on poor clusters, but 
large numbers of velocities are needed. 
 
     In Figure 4, we show velocity histograms for the 23 clusters with 
$\geq 10$ measurements from Table 1, as well as the four additional clusters 
listed in Table 3 (S49-132, N03-183, S49-128, \& MKW 2).  
With the exception of MKW 2, all 
plots are scaled to $C_{BI}\pm 2000~km~sec^{-1}$ to allow easy comparison 
of the dispersions between different clusters. 
We have used the ROSTAT
package (Beers, Flynn, \& Gebhardt 1990) to estimate measures of the normality of
the velocity histograms with respect to a Gaussian distribution.  
In dissipationless systems, dynamical evolution will be dominated 
by gravity.  N-body models show that over a relaxation time, 
galaxy-galaxy interactions 
will distribute the velocities in a Gaussian distribution.  Thus, we 
might expect that significant departures from a Gaussian may be indicative
of a non-equilibrium or non-virial dynamical state.  Ideally, however, one 
needs large numbers of velocities, and powerful 2-D and 3-D tests
to understand the true dynamical state of a cluster (see Pinkney \etal\ 1996). 
Velocity information is also limited to line-of-sight measurements, so 
the results are not unambiguous depending on the three dimensional spatial 
distribution of the galaxies ({\it i.e.\/} our viewing angle).  
The results, however, may be illuminating
with regards to the amount of velocity substructure present in our sample. 

     These statistical tests are very robust with respect to Poisson noise and 
low number statistics, 
however, erroneous results can be found due to 
outliers in the velocity distribution. 
We have used the $3\sigma$-clipping criteria as prescribed by Yahil \& 
Vidal (1977) to eliminate possible interlopers.  The inclusion of 
non-cluster members is particularly troublesome for poor clusters because
of the typically low number of member galaxies.  Even a few outliers
can severely bias the velocity dispersion and other scaling parameters.  
In addition to the
standard tests performed by ROSTAT (U-test, W-test, $B_1$-test, $B_2$-test,
$B_1-B_2$ omni-test, I-test, KS-stat, V-stat, $W^2$-stat, $U^2$-stat, 
U-stat, and $A^2$-stat), we also evaluate the Tail-index (TI) and 
Asymmetry-Index (AI) described in Bird \& Beers (1993) 
(see Beers \etal\ 1990, and
Pinkney \etal\ 1996 and references therein for a complete description
of these tests).  All of these statistics are sensitive to either
skewness or kurtosis (the third and fourth moments of the velocity 
distribution), and as such are useful measures of departures from a
true Gaussian distribution.  Pinkney \etal\ (1996) has shown that 
the $B_1$ and AI-stat are the most powerful discriminators of skewness. 
The W-test and TI-stat are the most sensitive to kurtosis. 
Many of the other tests do not separate these two effects,
but are useful for comparing the importance of one or the other effect 
as the cause of the non-normality.  

    Even after $3\sigma$ clipping, 55\% (15/27) of the clusters
show significant non-normality ($\geq 90\%$ significance) in at least one
of the statistics, which we regard as a conservative lower limit of the fraction 
of poor clusters in our sample with complex velocity distributions. 
For these clusters we have plotted velocity-coded
spatial (RA vs. DEC) diagrams in Figure 5.  These plots are useful
for locating 3-D substructure and galaxy clumps imbedded within 
the galaxy distribution.  In most cases, however, the velocity structure
in the clusters is quite complicated.  It is clear that in all 15 cases
there is no strong velocity/spatial segregation with respect to the
center or outskirts of the cluster.  This result is not specific to 
these 15 clusters, but appears to be true in general for our sample.
To quantify the 3-D nature of these clusters, 
we have also used a modified version of the 
Dressler-Shectman statistic (DS) (Dressler \& Shectman, 1988) 
to look for 3-d substructure.  Because of the small number of total 
galaxies in these clusters, we have used a modified version of the DS-test to 
use a window including the $\sqrt N$ nearest neighbors rather than
10 as suggested in the original test (Pinkney \etal\ 1996).  
We describe the results of the 1-D and DS 
statistics individually for all 15 clusters in the appendix.  

     As discussed in the previous section, 88\% of our sample is located
within or near a catalogued Zwicky cluster.  We have compared the 
velocity dispersions for the clusters divided by the open/medium-compact 
classification, 
and those not found within Zwicky's contours (isolated). 
We used the $S_{BI}$ ($\leq 0.5~Mpc$) values listed in Table 1. 
The median $S_{BI}$ are: isolated, $205\pm 66~km~sec^{-1}$; 
located within open clusters, $290\pm 47$; and medium-compact clusters, 
$329\pm 51$
(quoted errors are the error on the median). 
Within the errors these values are probably 
consistent, however, it is interesting that the median values steadily 
increase from isolated clusters to large, loose associations to more 
compact structures.  As described in the appendix, this situation is 
additionally complicated because of the location of what we define as 
a poor cluster within a more extended galaxy distribution.  Because of 
the surface-density requirement in selecting our poor clusters, it seems 
very likely that we are simply picking out local density maxima or 
clumps in a much larger gravitationally bound system.  Beers \etal\ (1995)
examined these trends with several AWM and MKW clusters and found
that many galaxy groups separated by as much as 3 Mpc in projection may 
be consistent with being gravitationally bound to one another (one very good 
example is the poor cluster pair MKW7/MKW8).  Only about 
12/63 of our clusters are found at the center of Zwicky's clusters. 
The remainder are located throughout the galaxy association, often 
embedded in a larger cluster with several other high surface-density
clumps.  

\section{Conclusions} 

     It is clear that the dynamics of poor clusters can be very complicated.
Even within the inner 1 Mpc, which we have investigated, over half of the 
clusters show evidence for non-equilibrium or non-virial conditions.  This
suggests that many of these clusters are in fact young, dynamically evolving
systems, which, in many cases, are simply a part of a much larger conglomerate
of galaxies.  

     Our poor clusters have very similar properties to other samples of 
poor clusters with a variety of selection criteria.  The presence of dominant
cluster galaxies or differences in the richness or compactness of the cluster
do not significantly change the observed velocity dispersions or other optical
properties.

As poor cluster samples and galaxy velocity databases grow larger, we may
gain a better appreciation for the complexity of hierarchical cluster 
formation.  We plan to use the statistical sample presented in this paper
for future studies, including an examination of the galaxy morphologies,
X-ray and radio properties, and the relationship to the richness and
the local dynamics.  

\acknowledgements

We acknowledge J. Pinkney for his help with the statistical analysis. 
We thank the Steward Observatory of the University of Arizona for 
generous allocation of observing time. We also thank T. Beers, the referee, for 
providing the ROSTAT analysis package, and for a careful reading of the
text. 
The Apache Point Observatory is maintained and operated by the 
Astrophysical Research Consortium (ARC).  
This research has made use of the NASA/IPAC Extragalactic Database (NED)
which is operated by the Jet Propulsion Laboratory and the California
Institute of Technology, under contract with the National Aeronautics \& 
Space Administration.  This research was supported by NSF Grant AST-9317596 
to J.O.B. and C.L. 

\clearpage 
\appendix 
\centerline{\bf Appendix: Comments on Individual Clusters}

\noindent 
{\bf S49-147} - The X-ray properties of this cluster are discussed in 
Price \etal\ (1991) and Dell'Antonio \etal\ (1994).  Burns \etal\ (1987)
report no radio identifications from their VLA survey.  The $B_1$, $B_2$,
$B_1-B_2$, and TI-tests all reported significant substructure.  The implied
skewness, however, is caused by the one galaxy at $6359~km~sec^{-1}$. 
This galaxy survived the $3\sigma$ clipping, but is responsible for 
the positive result.  S49-147 is located within the open cluster 
Zw0014.5+2315. 

\noindent 
{\bf AWM 1} - Also called N45-388, this cluster is discussed 
in detail by Beers \etal\ (1995),
who extend the search radius beyond 3 Mpc.  They found 3 separate groups
at distances between 1.9-2.7 Mpc which are consistent with being 
gravitationally bound to the primary cluster.  Within our 1 Mpc search
radius, we find mildy significant kurtosis to the right of the mean ($B_2$-
test).  The DS-test found significant clumping to the south and north-east 
of the dominant central galaxy.  Burns \etal\ (1987) found no radio 
galaxies in this cluster.  AWM 1 is located near the edge of the open 
cluster Zw0909.7+1814, which also includes N34-169 at the center.  

\noindent 
{\bf MKW 10} - Also called N67-312, this cluster is discussed 
by Beers \etal\ . In a more 
extended region around the cluster, they found 2 clumps at distances of
2.22 and 2.6 Mpc which may be bound to this cluster.  Within the inner
1 Mpc region, the velocity dispersion is quite small, but exhibits 
significant non-normality ($B_1$, $KS$, $W^2$, $U^2$, $A^2$, and AI-test)
which is attributed to skewness leftward of the mean velocity.  There is 
one compact radio source associated with an S0 galaxy.  MKW 10 is located
within the medium-compact cluster Zw1138.3+1024. 

\noindent 
{\bf N79-298B} - Only the $B_2$ and AI-tests were significant for this
cluster.  This is a result of the very sharp truncation of the 
velocity distribution at 4300 and 4900 $km~sec^{-1}$ (tails are too short). 
The DS-test found significant structure, which may be partly due to the
extreme elongation in the galaxy distribution.  This cluster is also 
called WP 19 (White 1978), and is located in a region near 
Abell 1367 (within the 
Coma supercluster) in the medium-compact cluster Zw1153.0+2522. 

\noindent 
{\bf N79-299B} - The 1-D non-normality is dominated by the two galaxies
between 6200-6450 $km~sec^{-1}$ which causes some degree of both 
skewness and kurtosis ($W$, $B_1$, $KS$, $W^2$, $U^2$, and $A^2$ tests). 
The 3-D distribution is very clumpy. However, the velocity mean and 
dispersion is very similar in both clumps which resulted in a negative
result for the DS-test.  There are two compact radio sources
associated with this cluster.  This cluster is also named WP 21, and is 
located within Zw1202.0+2028, a large open cluster which also includes
WP 20 (N79-299A, see below).  

\noindent 
{\bf N79-299A} - Only very moderate kurtosis is present in the velocity
distribution because of the sharp truncation to the right of the mean, 
and skewness from the slight asymmetry to the left.  The X-ray properties
of this cluster were discussed by Doe \etal\ (1995).  The spatial 
distribution is very elongated and clumpy, however the DS-test 
reported a negative result.  This cluster contains three compact radio
sources and a wide-angle tailed radio galaxy located at the mean 
cluster velocity.  N79-299A (or WP 20) is found within Zw1202.0+2028, in
the same large open cluster as N79-299B (WP 21).  

\noindent 
{\bf N67-330} - This is a very nearby, spiral-rich cluster.  The velocity 
distribution is strongly skewed, and several tests were significant 
($W$, $B_1$, $B_2$, $B_1-B_2$, $W^2$, $U^2$, $A^2$, and $U$).  
The average significance of 6 kurtosis tests was 97\% (90\% for skewness).  

\noindent 
{\bf N79-292} - The velocity distribution is strongly skewed to the right
of the mean ($W$, $B_1$, and AI tests had $>$95\% significance).  The
spatial distribution is very clumpy, however, the two primary groups 
have similar velocity fields.  The DS-test found a positive result (92\%) for 
the two galaxies to the south-west of the cluster.   N79-292 is not located
within a catalogued Zwicky cluster, but is near Zw1224.1+0914 (a medium-compact
cluster). 

\noindent 
{\bf MKW 11} - Beers \etal\ found two clumps $\sim$2.25 Mpc to the north 
with very similar velocities which are consistent with being bound to 
the primary cluster.  Within 1 Mpc, there is significant kurtosis to the 
left of the mean at the 90\% level ($B_2$ and $U$-tests).  There are 
two compact radio sources in this cluster.  Also called N79-296, MKW 11
is located within the open cluster Zw1327.3+1145.  There is another 
nearby Zwicky cluster in the immediate area as well.  

\noindent 
{\bf N79-286} - The X-ray properties of this cluster are discussed in 
Price \etal\ (1991).  All four velocities outside the primary peak 
in the histogram were clipped by the $3\sigma$-limit.  The velocity 
dispersion is quite low (146 $km~sec^{-1}$).  The $B_2$ and $U$-tests
found significant kurtosis (tails too short) with a slight skewness to
the left of the mean.  The 3-D distribution is very disperse with only
one tight grouping near the cluster center and no apparent substructure. 
This cluster is also identified with HCG 68.  

\noindent 
{\bf MKW 12} - Beers \etal\ note that this cluster is trimodal, with 
no significant large-scale structure outside the several groupings 
within the inner Mpc.  All but the $B_2$ and $B_1-B_2$ tests reported
significant non-normality.  This result is clearly due to the 
multimodal nature of the distribution.  Within the inner 1 Mpc, the DS-test
did not find significant 3-D clumping.  MKW 12 is also called N67-336, 
and is located at the center of Zw1400.4+0949, a medium-compact cluster. 

\noindent 
{\bf S49-132} - The X-ray and optical properties of this cluster are
discussed in Doe \etal\ (1995).  There are three radio galaxies within 
the central part of the cluster, each associated with a Zwicky galaxy
(two compact sources and one tailed source).  The velocity distribution
is very broad, with each of the radio/Zwicky galaxies separated by 
greater than  1000 $km~sec^{-1}$.  The very strong X-ray emission and 
the presence
of the tailed radio source argue against simple projection effects, however, 
the velocity distribution is very unusual for a poor cluster.  The $B_2$
$U$, and AI tests were significant at the 97\% level (tails too short 
for the large velocity dispersion).  The DS-test found significant 
3-D structure at the 94\% level.  
The elongation in the
galaxy distribution follows the distribution of the X-ray gas.  
One possible interpretation 
is that we are viewing a cluster which is accreting galaxies along a 
large-scale filament.  S49-132 is located within a larger galaxy isodensity
contour in the middle of a compact Zwicky cluster.  

\noindent 
{\bf N03-183} - This cluster is not part of the statistical sample in 
Burns \etal\ (1987) due to its low galactic latitude ($25^{\circ}$). 
It does, however, meet the surface-density enhancement requirement of 
the sample presented in this paper.  The cluster was observed as part of
a backup program with the MX.  The cluster is unusual in that we detected
7 emission-line galaxies from the MX spectra.  Most of the other clusters
contained one or possibly two such galaxies.  Positive signals were
found from the $W$, $B_1$, $U$, and AI-tests, indicating significant
asymmetries and skewness to the left of the mean, and moderate kurtosis 
(tails too short).  The DS-test did not find a positive signal for 3-D 
substructure.  

\noindent 
{\bf S49-128} - The X-ray properties of this cluster were discussed by
Doe \etal\ (1995).  The cluster contains a large double or Wide-angle
tailed radio source with a velocity of 6481 $km~sec^{-1}$, near the 
center of the primary velocity concentration.  The velocity distribution,
however, is very boxy, and strongly skewed to the left with slight kurtosis
to the right (short velocity tail) ($B_1$, $B_2$, $B_1-B_2$, $I$, $A^2$,
and AI-tests).  The single spatial outlier to the northwest falls within 
the $3\sigma$-limit, so was not clipped.  The distribution, however, 
is non-Gaussian even when this object is excluded.  S49-128 is found near
the north-edge of a large open Zwicky cluster. 

\noindent 
{\bf MKW2} - This optical properties of this cluster are discussed by 
Beers \etal\ (1995), the X-ray and radio by Doe \etal\ (1995).  The cluster
contains a wide-angle tailed radio source associated with a cD galaxy with
velocity 11260 $km~sec^{-1}$ (within the largest peak in the velocity
distribution).  While the velocity distribution appears bimodal (as also
noted by Beers \etal\ ) between $10000-12000~km~sec^{-1}$, the statistical
tests found only a weak non-Gaussian signature.  The 3 galaxies near 9000
and one galaxy near 13000 $km~sec^{-1}$ were $3\sigma$-clipped.  MKW 2 is
located near the edge of a medium-compact Zwicky cluster.

\clearpage 

\def\etal{{\it et al.\/}}
\begin{deluxetable}{rccccccccc}
\scriptsize
\tablewidth{0pt}
\tablenum{1}
\tablecaption{Poor Cluster Sample} 
\tablehead{
\colhead{Cluster}           & \colhead{RA(2000)}      &
\colhead{DEC(2000)}          & \colhead{z}  &
\colhead{N}       & \colhead{$S_{BI}$} & 
\colhead{N}          & \colhead{$S_{BI}$}    &
\colhead{Ref} &\colhead{Other Name} \\
\colhead{}      &  \colhead{}  &
\colhead{}      & \colhead{}   &
\colhead{$\leq 0.5~Mpc$}  & \colhead{$\leq 0.5~Mpc$} &
\colhead{$\leq 1~Mpc$} & \colhead{$\leq 1~Mpc$}} 
\startdata
S34--115& 00 18 18.7& 30 02 41 & 0.0225 & 10 & 474$^{+168}_{-93}$ & 12 & \nodata & NED  \nl
S49--147& 00 21 30.1& 22 26 39 & 0.0191 & 8 & 289$^{+172}_{-84}$ & 12 & 233$^{+141}_{-43}$ & NED  \nl
S34--111& 01 07 27.7& 32 23 59 & 0.0173 & 29 & 466$^{+55}_{-41}$ & 47 & 486$^{+53}_{-37}$ & MX   \nl
S49--140& 01 56 22.9& 05 37 37 & 0.0179 & 10 & 205$^{+59}_{-21}$ & 13 & \nodata & MX  \nl
S49--141& 02 01 47.8& 08 28 25 & 0.0264 & 4  & \nodata & 4 & \nodata & APO \nl
S49--145& 02 07 34.9& 02 08 14 & 0.0227 & 6 & 519$^{+41}_{-30}$ & 7 & \nodata & NED  \nl
S49--142& 03 20 44.7&--01 02 15 & 0.0211 & 6 & 69$^{+2}_{-2}$ & 7 & \nodata & NED  \nl
N45--342& 09 10 10.9& 50 24 47 & 0.0165 & 3 & \nodata & 3 & \nodata & NED  \nl
N34--169& 09 16 11.8& 17 39 28 & 0.0292 & 3 & \nodata & 3 & \nodata & APO/NED  \nl
N45--388& 09 17 31.6& 19 51 24 & 0.0291 & 7 & 584$^{+238}_{-193}$ & 20 & 608$^{+94}_{-53}$ & BKBH & AWM 1 \nl
*N45--366& 09 23 39.3& 22 21 07 & 0.0316 & 6 & 888$^{+226}_{-65}$ & 6 & \nodata & APO/NED  \nl
N45--384& 09 27 51.8& 29 59 56 & 0.0266 & 7 & 238$^{+125}_{-14}$ & 11 & 185$^{+84}_{-40}$ & NED  \nl
N34--170& 09 42 24.6& 04 16 16 & 0.0292 & 2 & \nodata & 2 & \nodata & APO \nl
N34--172& 10 00 32.1&--02 57 27 & 0.0207 & 8 & 337$^{+135}_{-45}$ & 9 & \nodata & BKBH & MKW 1 \nl  
N56--393& 10 13 52.0& 38 40 07 & 0.0221 & 6 & 422$^{+99}_{-24}$ & 6 & \nodata & NED  \nl
*N56--387& 10 25 18.2& 17 08 44 & 0.0025 & 3 & \nodata & 3 & \nodata & NED \nl 
*N56--367& 10 27 05.5& 16 04 40 & 0.0331 & 2 & \nodata & 2 & \nodata & APO \nl
*N56--388& 10 41 24.4& 06 10 17 & 0.0310 & 3 & \nodata & 3 & \nodata & NED \nl
*N45--371& 11 20 12.8& 72 50 35 & 0.0373 & 2 & \nodata & 2 & \nodata & APO/NED \nl
*N56--396& 11 21 28.3& 02 52 33 & 0.0496 & 3 & \nodata & 3 & \nodata & APO \nl
N67--311& 11 22 26.8& 24 17 33 & 0.0264 & 6 & 149$^{+186}_{-53}$ & 6 & \nodata & RDGH & HCG 51 \nl
N67--322& 11 28 35.1& 09 05 29 & 0.0210 & 6 & 45$^{+18}_{-16}$ & 6 & \nodata & NED   \nl
*N56--359& 11 36 21.3&--02 50 37 & 0.0454 & 2 & \nodata & 2 & \nodata & APO \nl
*N79--278& 11 37 54.4& 21 58 23 & 0.0306 & 8 & 344$^{+57}_{-14}$ & 8 & \nodata & RDGH & HCG 57 \nl
N67--312& 11 42 04.6& 10 18 20 & 0.0206 & 9 & 177$^{+85}_{-46}$ & 11 & \nodata & BKBH & MKW 10 \nl
*N79--290& 11 44 12.9& 33 40 20 & 0.0325 & 4 & \nodata & 5 & 236$^{+67}_{-6}$ & APO/NED \nl
N56--371& 11 45 03.5&--01 39 38 & 0.0276 & 3 & \nodata & 3 & \nodata & APO/NED \nl
*N79--280& 11 46 18.5& 33 09 19 & 0.0325 & 4 & \nodata & 4 & \nodata & NED   \nl
N67--300& 11 48 22.5& 12 43 19 & 0.0129 & 4 & \nodata & 5 & 446$^{+226}_{-256}$ & RDGH & HCG 59 \nl
N56--392& 11 49 38.9&--03 31 35 & 0.0272 & 8 & 106$^{+167}_{-58}$ & 8 & \nodata & APO/BKBH & MKW 3 \nl
N79--298& 11 57 52.3&25 10 18 & 0.0153 & 12 & 185$^{+30}_{-32}$ & 13 & \nodata & NED \nl
N79--299B& 12 04 09.5&20 13 18 & 0.0235 & 5 & 232$^{+159}_{-3}$ & 14 & 384$^{+151}_{-110}$ & NED \nl
N67--335& 12 04 21.7& 01 50 19 & 0.0204 & 22 & 607$^{+135}_{-52}$ & 31 & 568$^{+85}_{-70}$ & BKBH & MKW 4 \nl
N79--299A& 12 05 51.2&20 32 19 & 0.0235 & 9 & 495$^{+130}_{-72}$ & 15 & 419$^{+84}_{-49}$ & NED \nl
N79--282& 12 12 25.9& 29 08 20 & 0.0132 & 5 & 85$^{+37}_{-4}$ & 10 & 89$^{+19}_{-11}$ & NED & HCG 61 \nl
N79--268& 12 19 42.9& 28 47 22 & 0.0253 & 4 & \nodata & 4 & \nodata & NED   \nl
N79--283& 12 19 54.8& 28 25 21 & 0.0259 & 2 & \nodata & 2 & \nodata & APO \nl
*N67--330& 12 20 02.3& 05 20 24 & 0.0068 & 32 & 413$^{+76}_{-40}$ & 55 & 485$^{+63}_{-41}$ & NED   \nl
N79--292& 12 24 14.7& 09 20 24 & 0.0235 & 7 & 445$^{+88}_{-13}$ & 18 & 558$^{+147}_{-95}$ & NED   \nl
N79--284& 12 35 58.4& 26 58 29 & 0.0246 & 4 & \nodata & 7 & 598$^{+277}_{-69}$ & NED   \nl
*N67--333& 13 04 25.3& 07 54 54 & 0.0449 & 2 & \nodata & 2 & \nodata & APO \nl 
N67--323& 13 05 26.5& 53 33 56 & 0.0289 & 4 & \nodata & 4 & \nodata & APO \nl 
N67--317& 13 13 49.0& 06 57 09 & 0.0217 & 2 & \nodata & 2 & \nodata & APO \nl 
N79--270& 13 17 19.3& 20 37 11 & 0.0226 & 4 & \nodata & 4 & \nodata & NED \nl 
N79--296& 13 29 22.3& 11 47 31 & 0.0232 & 8 & 424$^{+146}_{-21}$ & 14 & 384$^{+70}_{-42}$ & BKBH & MKW 11 \nl 
N67--329& 13 32 36.4& 07 20 36 & 0.0231 & 5 & 155$^{+54}_{-51}$ & 5 & \nodata & APO/NED \nl 
N67--318& 13 52 19.0& 02 20 19 & 0.0235 & 2 & \nodata & 2 & \nodata & APO \nl 
*N79--286& 13 53 31.0& 40 16 17 & 0.0085 & 14 & 145$^{+173}_{-18}$ & 27 & 183$^{+77}_{-36}$ & NED & HCG 68 \nl 
N79--297& 13 55 24.7& 25 03 20 & 0.0293 & 7 & 178$^{+103}_{-42}$ & 7 & \nodata & RDGH & HCG 69 \nl 
*N79--276& 13 56 22.3& 28 31 23 & 0.0358 & 3 & \nodata & 3 & \nodata & NED \nl 
N67--336& 14 03 04.0& 09 26 35 & 0.0196 & 18 & 625$^{+153}_{-118}$ & 31 & 721$^{+176}_{-162}$ & BKBH & MKW 12 \nl 
N67--325& 14 09 58.5& 17 32 51 & 0.0171 & 5 & 320$^{+66}_{-54}$ & 6 & \nodata & NED \nl 
*N56--361& 14 15 12.6& 04 52 04 & 0.0559 & 2 & \nodata & 2 & \nodata & APO/NED \nl 
N67--326& 14 28 14.1& 25 50 38 & 0.0153 & 8 & 299$^{+83}_{-41}$ & 14 & 289$^{+50}_{-39}$ & BKBH & AWM 3 \nl 
N67--309& 14 28 31.6& 11 22 38 & 0.0265 & 6 & 292$^{+97}_{-27}$ & 8 & \nodata & APO/NED \nl 
N56--394& 14 34 00.9& 03 44 53 & 0.0289 & 8 & 663$^{+387}_{-440}$ & 16 & 573$^{+363}_{-172}$ & BKBH & MKW 7 \nl
N56--395& 14 40 43.2& 03 27 12 & 0.0272 & 7 & 329$^{+110}_{-97}$ & 15 & 422$^{+99}_{-53}$ & BKBH & MKW 8 \nl 
N56--381& 14 47 00.4& 11 35 29 & 0.0295 & 5 & 265$^{+22}_{-22}$ & 6 & \nodata & NED \nl 
*N67--310& 14 48 06.2& 31 44 32 & 0.0453 & 5 & 111$^{+28}_{-4}$ & 5 & \nodata & APO/NED \nl
*N56--365& 14 58 26.7& 48 27 59 & 0.0370 & 2 & \nodata & 2 & \nodata & APO \nl
*N45--381& 15 13 11.6& 04 28 50 & 0.0360 & 4 & \nodata & 4 & \nodata & APO/NED \nl 
N56--374& 15 36 30.3& 43 31 08 & 0.0189 & 5 & 55$^{+16}_{-2}$ & 6 & \nodata & NED \nl
*N45--361& 15 49 36.8& 42 01 56 & 0.0355 & 3 & \nodata & 3 & \nodata & APO/NED \nl
*N45--363& 15 57 46.9& 16 16 27 & 0.0354 & 3 & \nodata & 3 & \nodata & NED \nl
*N45--389& 16 17 39.2& 35 05 45 & 0.0301 & 9 & 239$^{+87}_{-17}$ & 11 & \nodata & NED \nl
N34--171& 16 41 35.4& 57 50 21 & 0.0176 & 3 & \nodata & 3 & \nodata & APO \nl 
N34--175& 17 15 21.4& 57 22 43 & 0.0283 & 7 & 589$^{+440}_{-31}$ & \nodata & \nodata & APO/NED \nl
N34--173& 17 55 24.9& 62 36 40 & 0.0266 & 4 & \nodata & 4 & \nodata & NED \nl
S49--146& 22 50 17.7& 11 35 53 & 0.0250 & 6 & 482$^{+102}_{-54}$ & 10 & 617$^{+132}_{-110}$ & NED \nl
*S49--143& 23 01 48.9& 15 58 09 & 0.0077 & 8 & 153$^{+55}_{-16}$ & 11 & 136$^{+37}_{-18}$ & NED \nl
S49--144& 23 16 24.2& 15 51 22 & 0.0146 & 2 & \nodata & 2 & \nodata & NED \nl
\tablerefs{(BKBH) Beers, T.C. \etal\ 1995; (RDGH) Ramella, M. \etal\ 1994; (APO)
 This paper; (MX) This paper; (NED) Found through search of the NED Extragalactic database.} 
\tablenotetext{*}{Cluster is not part of statistical sample for $0.01<z<0.03$.}
\enddata
\end{deluxetable}

\pagestyle{empty}

\begin{deluxetable}{llccrccr}
\small
\tablewidth{33pc}
\tablenum{2}
\tablecaption{ARC 3.5-m Velocity Measurements} 
\tablehead{
\colhead{Cluster}         &  \colhead{Galaxy} &
\colhead{RA(2000)}        &  \colhead{DEC(2000)} &
\colhead{$V_H$}           &  \colhead{Error} &
\colhead{Hubble Type}}     
\startdata
S49-141& NGC 0791 & 02 01 44.32 & 08 30 00.1 & 7905 &  75 & E \nl 
& UGC 01513 & 02 01 47.23 & 08 28 35.9 & 7547 &  100 & S/I \nl 
N34-169& 0915+176 & 09 15 56.26 & 17 38 33.8 & 9046 &  200 & S \nl 
N45-366& UGC 04991 \#1 & 09 23 27.17 & 22 19 33.6 & 9558 &  95 & E \nl
& UGC 04991 \#2 & 09 23 25.99 & 22 19 01.2 & 9293 &  95 & E \nl 
& UGC 04991 \#3 & 09 23 24.18 & 22 18 50.4 & 10261 &  105 & E \nl 
& 0923+223a & 09 23 31.32 & 22 18 30.6 & 9117 &  77 & S \nl 
& *0923+223b & 09 23 29.97 & 22 19 38.9 & 8219 &  100 & E \nl 
N34-170& UGC 05182 & 09 42 25.04 & 04 16 57.2 & 8631 &  48 & E \nl 
& *CGCG 035-032 & 09 42 13.38 & 04 16 25.2 & 8469 &  50 & S \nl 
N56-367 & CGCG 094-058 & 10 27 05.47 & 16 00 41.4 & 9653 &  89 & S \nl
& CGCG 094-062 & 10 27 14.74 & 16 02 56.8 & 10171 &  73 & E \nl 
N45-371 & 1121+728 & 11 21 28.77 & 72 48 29.6 & 11194 &  115 & S \nl
N56-396 & CGCG 039-135 & 11 21 32.56 & 02 53 13.8 & 14616 &  56 & E \nl
& NGC 3647     & 11 21 38.58 & 02 53 30.0 & 14714 &  80 & E \nl
& CGCG 039-140 & 11 21 35.30 & 02 53 37.5 & 15256 &  82 & E \nl
N56-359 & CGCG 012-037 & 11 36 19.29 & -02 50 50.1 & 13812 &  150 & S \nl
& CGCG 012-036 & 11 36 14.01 & -02 52 13.8 & 13418 &  160 & S \nl 
N79-290& *CGCG 186-029N1 & 11 44 03.35 & 33 32 06.2 & 9501 &  100 & S \nl 
& CGCG 186-029N2 & 11 44 04.36 & 33 32 33.8 & 9230 &  100 & S \nl 
N56-371 & CGCG 012-076 & 11 45 15.03 & -01 42 11.0 & 8147 &   87 & E \nl 
& CGCG 012-074 & 11 45 06.38 & -01 40 09.3 & 8220 &  107 & S \nl 
N56-392 & CGCG 012-098 & 11 49 39.72 & -03 31 47.2 & 8210 &  120 & S \nl
& CGCG 012-095 & 11 49 35.44 & -03 29 18.0 & 8215 &  78 & E \nl
& 1149-034 & 11 49 32.28 & -03 28 34.2 & 8141 &  80 & I \nl
N79-283 & CGCG 158-074 & 12 19 54.80 & 28 23 21.2 & 7594 &  126 & S \nl
& CGCG 158-078 & 12 20 10.21 & 28 23 19.9 & 7912 &  160 & E  \nl
N67-333 & CGCG 043-128 & 13 04 14.44 & 07 54 13.4 & 13646 &  98 & E \nl
& CGCG 043-127 & 13 04 07.20 & 07 54 55.3 & 13326 &  140 & E \nl 
N67-323 & NGC 4967 & 13 05 36.42 & 53 33 51.9 & 8812 &  145 & E \nl
& MRK 0239 & 13 05 25.82 & 53 35 30.5 & 11331 &  98 & E \nl
& 1305+536 & 13 05 19.61 & 53 35 39.5 & 15636 &  112 & I \nl
& 1305+535 & 13 05 14.38 & 53 34 12.7 & 9126 &  150 & S \nl
& 1304+536 & 13 04 58.64 & 53 36 34.5 & 26880 &  140 & S \nl
& 1304+535 & 13 04 56.87 & 53 35 29.9 & 9405 &  129 & E \nl
N67-317 & CGCG 044-036 & 13 13 55.17 & 06 57 07.7 & 6375 &  87 & S \nl
& CGCG 044-033 & 13 13 44.77 & 06 59 22.6 & 6620 &  140 & S \nl
N67-329 & CGCG 045-011 & 13 32 54.65 & 07 19 39.4 & 7099 &  78 & S \nl
& NGC 5209     & 13 32 42.55 & 07 19 37.7 & 7042 &  74 & E \nl
& CGCG 045-008 & 13 32 32.31 & 07 17 38.9 & 6821 &  150 & S \nl
N67-318 & CGCG 045-124 & 13 52 22.76 & 02 20 43.0 & 6981 &  150 & S/E \nl
& CGCG 045-122 & 13 52 14.53 & 02 21 33.2 & 11149 &  200 & E \nl
N56-361 & 1415+049 & 14 15 16.77 & 04 54 28.3 & 16819 &  98 & S \nl
& CGCG 046-076 & 14 15 12.07 & 04 52 19.6 & 13796 &  150 & S \nl
& CGCG 046-073 & 14 14 54.24 & 04 54 15.7 & 22396 &  160 & S \nl 
N67-309 & CGCG 075-048 & 14 28 40.83 & 11 22 05.0 & 8184 &  130 & S \nl
& *CGCG 075-043 & 14 28 20.68 & 11 21 33.9 & 7766 &  100 & S \nl
& CGCG 075-044 & 14 28 22.43 & 11 25 05.0 & 7914 &   63 & E \nl
N67-310 & CGCG 164-027 & 14 48 08.35 & 31 42 36.7 & 9888 &  117 & E \nl
& CGCG 164-028 & 14 48 07.35 & 31 44 16.7 & 13567 &  128 & E \nl
& IC 4508 (1)  & 14 47 50.77 & 31 45 53.2 & 13513 &  98 & E \nl
& IC 4508 (2)  & 14 47 47.90 & 31 46 51.3 & 13700 &  107 & E \nl
& *1448+317     & 14 48 12.89 & 31 44 33.2 & 13639 &      & I \nl
N56-365 & CGCG 248-035 & 14 58 32.27 & 48 29 04.6 & 10993 &  75 & E \nl
& CGCG 248-034 & 14 58 27.76 & 48 29 32.6 & 11210 &  74 & E \nl
N45-381 & 1513+044 & 15 13 28.14 & 04 27 56.4 & 15964 &  100 & S \nl
& CGCG 049-054 & 15 13 18.60 & 04 28 38.9 & 11484 &  102 & E \nl
& CGCG 049-049 & 15 13 10.43 & 04 28 55.5 & 10980 &  85 & E \nl
& CGCG 049-045 & 15 13 08.24 & 04 29 49.3 & 9829 &  150 & S \nl 
N45-361 & CGCG 222-052 & 15 49 38.28 & 42 02 03.5 & 10076 &  96 & E \nl 
N34-171& NGC 6211 & 16 41 27.83 & 57 47 01.4 & 5248 &  69 & E \nl
& NGC 6213 & 16 41 37.22 & 57 48 53.8 & 5377 &  46 & S \nl
& UGC 10516 & 16 41 44.87 & 57 50 55.9 & 5204 &  78 & S \nl
N34-175& IC 1252 & 17 15 50.11 & 57 22 01.1 & 8996 &  108 & S \nl 
& NPM1G+57.0229 & 17 15 47.08 & 57 18 08.0 & 7822 &  90 & S \nl 
& 1715+573a & 17 15 29.18 & 57 21 00.1 & 8430 &  150 & S/S0 \nl
& NGC 6346 & 17 15 24.26 & 57 19 22.5 & 8739 &  73 & E \nl 
& 1715+573b & 17 15 16.58 & 57 19 47.5 & 7732 &  91 & S/S0 \nl 
& 1715+573c & 17 15 08.32 & 57 18 39.7 & 9280 &  63 & S \nl 
\tablenotetext{}{$^*$ Emission-Line Redshift} 
\enddata
\end{deluxetable}

\pagestyle{empty}

\begin{deluxetable}{lcccrccr}
\small
\tablewidth{33pc}
\tablenum{3}
\tablecaption{Steward Observatory 2.3-m MX Velocity Measurements} 
\tablehead{
\colhead{Cluster}         &  \colhead{Galaxy} &
\colhead{RA(2000)}        &  \colhead{DEC(2000)} &
\colhead{$V_H$}           &  \colhead{error}}
\startdata
S49--132 & &  23:10:31.81 & +07:34:17.8 \nl 
& 201 & 23:10:42.30 & +07:34:04.1 & 13204 & 45 \nl
& 203 & 23:10:40.60 & +07:23:16.6 & 10979 & 51 \nl
& 204 & 23:11:33.75 & +07:31:49.6 & 11619 & 44 \nl
& 208 & 23:11:07.85 & +07:24:24.1 & 12149 & 61 \nl
& 210 & 23:10:38.28 & +07:33:54.9 & 12967 & 57 \nl
& 211 & 23:10:37.36 & +07:32:48.7 & 12352 & 38 \nl
& 212 & 23:10:33.43 & +07:32:19.5 & 12208 & 48 \nl
& 215 & 23:10:49.98 & +07:27:20.5 & 12757 & 82 \nl
& 220 & 23:10:45.52 & +07:26:14.2 & 11777 & 88 \nl
& 226 & 23:10:20.74 & +07:22:14.8 & 12245 & 40 \nl
& 227 & 23:10:18.12 & +07:32:42.2 & 10818 & 50 \nl
& 228 & 23:09:53.38 & +07:29:31.6 & 13752 & 49 \nl
& 229 & 23:09:09.44 & +07:30:26.1 & 11095 & 40 \nl
& *230 & 23:09:03.92 & +07:31:04.3 & 12314 & 56 \nl
& 231 & 23:09:59.04 & +07:20:32.9 & 11734 & 46 \nl
& 234 & 23:10:14.43 & +07:27:09.3 & 13355 & 40 \nl
& 236 & 23:10:20.81 & +07:32:36.2 & 11225 & 38 \nl
& 237 & 23:10:27.94 & +07:32:46.1 & 12771 & 50 \nl
& 238 & 23:10:03.08 & +07:34:06.7 & 11241 & 39 \nl
& 239 & 23:10:18.51 & +07:34:17.8 & 13573 & 108 \nl
& 244 & 23:10:09.46 & +07:16:19.4 & 13401 & 45 \nl
& 246 & 23:09:54.33 & +07:22:16.5 & 11907 & 47 \nl
& 250 & 23:10:30.43 & +07:35:20.8 & 12651 & 39 \nl
& 252 & 23:10:48.48 & +07:35:12.0 & 12620 & 46 \nl
& 254 & 23:10:58.94 & +07:39:38.8 & 12838 & 41 \nl
& 256 & 23:11:35.66 & +07:44:35.0 & 12724 & 40 \nl
& 262 & 23:10:22.39 & +07:34:51.0 & 11852 & 38 \nl
& 263 & 23:10:25.11 & +07:34:42.0 & 11048 & 37 \nl
& 264 & 23:10:26.86 & +07:35:24.9 & 11273 & 48 \nl
& 265 & 23:10:25.88 & +07:37:39.0 & 11274 & 45 \nl
& 266 & 23:10:14.73 & +07:39:08.6 & 10858 & 108 \nl
& 269 & 23:10:06.28 & +07:46:02.6 & 11775 & 41 \nl
& 271 & 23:10:22.53 & +07:53:25.1 & 15107 & 58 \nl
& *272 & 23:10:00.06 & +07:35:42.9 & 11073 & 75 \nl
\nl
N56--394 & & 14:34:00.94 & +03:44:53.4 & \nl 
& 250 & 14:33:33.52 & +03:16:28.4 & 9190 & 122 \nl 
& 256 & 14:33:59.11 & +03:46:40.7 & 8722 & 25 \nl 
& 257 & 14:33:58.08 & +03:44:47.9 & 10037 & 19 \nl 
& 258 & 14:34:03.94 & +03:44:51.2 & 7366 & 18 \nl 
& 259 & 14:34:01.44 & +03:47:50.4 & 8779 & 24 \nl 
& 261 & 14:33:55.29 & +03:50:02.6 & 8739 & 24 \nl 
& 263 & 14:33:30.74 & +03:41:08.3 & 8418 & 50 \nl 
& 264 & 14:33:51.41 & +03:40:45.8 & 8702 & 27 \nl 
& 266 & 14:34:50.57 & +03:38:41.8 & 8537 & 82 \nl 
& 267 & 14:34:08.23 & +03:54:05.4 & 8616 & 22 \nl 
& 268 & 14:33:17.49 & +03:54:11.9 & 8942 & 57 \nl 
& 269 & 14:33:48.30 & +03:57:24.4 & 8810 & 30 \nl 
& 271 & 14:33:34.49 & +03:55:52.4 & 25531 & 40 \nl 
& 282 & 14:33:38.23 & +03:41:46.0 & 9797 & 35 \nl 
& 283 & 14:33:41.87 & +03:41:38.0 & 26220 & 73 \nl 
& 284 & 14:33:11.88 & +03:48:20.9 & 25784 & 60 \nl 
& 286 & 14:33:09.26 & +03:34:29.3 & 9227 & 90 \nl 
& 288 & 14:34:05.29 & +03:39:26.9 & 8792 & 131 \nl 
& 329 & 14:33:00.51 & +03:47:25.7 & 8864 & 45 \nl 
& *331 & 14:32:41.70 & +03:46:30.9 & 8406 & 64 \nl 
& 337 & 14:32:49.92 & +03:40:23.8 & 45093 & 97 \nl 
& 347 & 14:32:35.39 & +03:16:51.8 & 13853 & 125 \nl 
& *352 & 14:31:27.52 & +03:25:03.9 & 16064 & 122 \nl 
& 360 & 14:32:21.32 & +03:24:49.7 & 12834 & 78 \nl
& 361 & 14:32:09.13 & +03:25:52.3 & 32321 & 88 \nl 
& *363 & 14:31:53.07 & +03:22:47.8 & 1497 & 50 \nl 
& *366 & 14:32:29.21 & +03:27:24.3 & 8556 & 50 \nl 
\nl
N03--183 & & 07:10:50.36 & +50:10:03.1& \nl 
& 214 & 07:12:25.68 & +50:39:37.1 & 6480 & 59 \nl 
& 227 & 07:12:43.63 & +50:03:41.7 & 5487 & 34 \nl 
& 235 & 07:12:04.49 & +50:35:31.5 & 5132 & 48 \nl 
& 236 & 07:12:03.86 & +50:27:50.6 & 5818 & 95 \nl 
& 237 & 07:11:29.04 & +50:40:01.7 & 14797 & 99 \nl 
& 238 & 07:11:22.59 & +50:31:49.2 & 6206 & 68 \nl 
& *240 & 07:10:35.10 & +50:47:40.9 & 5816 & 50 \nl 
& 241 & 07:09:52.71 & +50:27:00.3 & 5686 & 24 \nl 
& 247 & 07:11:59.72 & +50:15:04.5 & 5777 & 103 \nl 
& 249 & 07:11:33.59 & +50:14:53.0 & 6035 & 24 \nl 
& 253 & 07:11:04.66 & +50:08:11.7 & 4846 & 39 \nl 
& 255 & 07:10:49.07 & +50:14:48.3 & 23195 & 122 \nl 
& 257 & 07:10:38.62 & +50:10:37.3 & 5648 & 33 \nl 
& 258 & 07:10:44.98 & +50:04:52.4 & 5713 & 28 \nl 
& 259 & 07:10:34.10 & +50:07:07.0 & 6490 & 23 \nl 
& 261 & 07:10:25.87 & +49:56:29.9 & 18076 & 122 \nl 
& 262 & 07:10:02.59 & +50:17:41.2 & 18411 & 43 \nl 
& 263 & 07:09:42.51 & +50:03:55.9 & 19451 & 122 \nl 
& 265 & 07:09:28.36 & +50:09:08.4 & 4887 & 34 \nl 
& 266 & 07:11:52.00 & +49:49:33.3 & 18621 & 36 \nl 
& 267 & 07:11:41.78 & +49:51:43.5 & 6164 & 22 \nl 
& 268 & 07:11:37.50 & +49:45:56.5 & 18589 & 47 \nl 
& 269 & 07:11:28.52 & +49:33:37.8 & 5865 & 122 \nl 
& 270 & 07:11:16.31 & +49:48:15.0 & 18398 & 40 \nl 
& 271 & 07:11:08.92 & +49:53:59.3 & 6296 & 77 \nl 
& 273 & 07:10:53.95 & +49:54:11.1 & 6393 & 19 \nl 
& 274 & 07:10:34.75 & +49:52:22.2 & 6174 & 26 \nl 
& 275 & 07:09:51.01 & +49:52:03.4 & 5427 & 50 \nl 
& 276 & 07:09:30.18 & +49:39:11.8 & 18580 & 50 \nl 
& 278 & 07:08:38.73 & +50:35:01.0 & 13336 & 119 \nl 
& *279 & 07:08:34.23 & +50:37:51.4 & 6164 & 50 \nl 
& 286 & 07:09:20.48 & +50:02:48.9 & 18005 & 45 \nl 
& *300 & 07:09:17.82 & +49:52:16.1 & 4919 & 50 \nl 
& *301 & 07:09:15.87 & +49:49:12.9 & 4710 & 50 \nl 
\nl
S49--140 & & 01:56:22.89 & +05:37:37.3 & \nl 
& 232 & 01:58:45.97 & +05:31:30.9 & 6715 & 122 \nl 
& 234 & 01:58:46.81 & +05:47:21.1 & 28341 & 56 \nl 
& *253 & 01:57:22.60 & +05:42:42.6 & 4263 & 122 \nl 
& 261 & 01:58:07.32 & +05:20:16.7 & 31936 & 122 \nl 
& 300 & 01:57:03.63 & +05:27:42.9 & 8454 & 113 \nl 
& 304 & 01:56:52.10 & +05:46:29.0 & 5378 & 40 \nl 
& 309 & 01:56:36.80 & +05:48:14.7 & 5655 & 81 \nl 
& 311 & 01:56:42.87 & +05:41:16.4 & 5154 & 53 \nl 
& 322 & 01:56:19.03 & +05:39:07.3 & 6421 & 76 \nl 
& 326 & 01:56:11.99 & +05:35:18.6 & 5253 & 42 \nl 
& 326 & 01:56:11.99 & +05:35:18.6 & 5349 & 53 \nl 
& 355 & 01:56:02.38 & +05:00:04.0 & 9930 & 122 \nl 
& 357 & 01:55:27.56 & +05:21:31.9 & 36395 & 122 \nl 
& 379 & 01:55:19.26 & +05:27:58.6 & 21073 & 122 \nl 
& 385 & 01:54:42.42 & +05:25:34.4 & 47490 & 93 \nl 
& 398 & 01:55:06.94 & +05:22:24.7 & 5551 & 95 \nl 
& 399 & 01:55:02.54 & +05:08:19.0 & 25329 & 108 \nl 
& 402 & 01:54:41.82 & +05:05:41.0 & 47620 & 103 \nl 
& 405 & 01:53:55.53 & +05:18:37.1 & 43495 & 113 \nl
\nl
S34-111 & & 01:07:27.71 & +32:23:59.7 & \nl 
& 241 & 01:07:46.16 & +32:41:34.2 & 5434 & 31 \nl 
& 244 & 01:09:42.48 & +32:27:04.9 & 5081 & 56 \nl 
& 259 & 01:08:53.69 & +32:30:51.9 & 5051 & 35 \nl 
& 261 & 01:08:15.85 & +32:29:56.3 & 5661 & 75 \nl 
& 265 & 01:08:12.94 & +32:27:12.4 & 4846 & 26 \nl 
& 271 & 01:07:24.98 & +32:24:45.1 & 5212 & 28 \nl 
& 274 & 01:09:59.37 & +32:22:05.8 & 5247 & 27 \nl 
& 286 & 01:09:33.28 & +32:10:23.2 & 5606 & 31 \nl 
& 288 & 01:09:14.03 & +32:09:04.7 & 4757 & 47 \nl 
& 293 & 01:08:55.26 & +32:19:30.4 & 4735 & 59 \nl 
& 295 & 01:08:58.94 & +32:14:25.8 & 4514 & 93 \nl 
& 296 & 01:08:52.33 & +32:05:59.0 & 10617 & 70 \nl 
& 298 & 01:08:18.79 & +32:13:35.2 & 5315 & 92 \nl 
& 299 & 01:08:43.92 & +32:06:06.4 & 5718 & 87 \nl 
& 300 & 01:08:17.26 & +32:05:24.1 & 4652 & 69 \nl 
& 304 & 01:07:47.17 & +32:18:35.6 & 5534 & 25 \nl 
& 305 & 01:07:31.30 & +32:21:42.7 & 5709 & 31 \nl 
& 306 & 01:07:33.06 & +32:23:27.5 & 4824 & 33 \nl 
& 311 & 01:07:32.60 & +32:05:34.0 & 14851 & 35 \nl 
& 318 & 01:09:13.50 & +31:58:45.8 & 5255 & 25 \nl 
& 357 & 01:07:05.78 & +32:47:42.2 & 5045 & 35 \nl 
& 358 & 01:06:51.44 & +32:42:04.4 & 5945 & 48 \nl 
& 359 & 01:06:57.05 & +32:37:20.3 & 19986 & 32 \nl 
& 360 & 01:07:15.68 & +32:31:12.9 & 5492 & 31 \nl 
& 361 & 01:07:17.60 & +32:28:58.8 & 4465 & 34 \nl 
& 364 & 01:06:53.76 & +32:34:42.7 & 5558 & 116 \nl 
& 375 & 01:05:56.00 & +32:24:43.7 & 4830 & 29 \nl 
& 391 & 01:07:27.22 & +32:19:11.5 & 5156 & 27 \nl 
& 392 & 01:07:25.01 & +32:17:33.8 & 4313 & 29 \nl 
& 395 & 01:07:11.55 & +32:10:13.4 & 4821 & 52 \nl 
& 401 & 01:07:05.92 & +32:20:53.0 & 5965 & 28 \nl 
& 402 & 01:06:58.21 & +32:18:30.4 & 5581 & 25 \nl 
\nl 
S49-128 & & 02:41:04.23 & +08:43:48.5 & \nl 
& 248 & 02:41:55.63 & +08:59:48.3 & 6103 & 90 \nl 
& 250 & 02:42:07.22 & +08:57:08.0 & 21256 & 45 \nl 
& 270 & 02:42:45.29 & +08:53:14.6 & 21780 & 64 \nl 
& 271 & 02:42:44.55 & +08:50:28.7 & 21178 & 57 \nl 
& 276 & 02:42:26.84 & +08:49:52.1 & 21296 & 46 \nl 
& 281 & 02:41:50.03 & +08:52:45.0 & 6625 & 35 \nl 
& 289 & 02:41:16.28 & +08:45:38.1 & 6421 & 90 \nl 
& 290 & 02:41:06.12 & +08:44:16.5 & 6365 & 37 \nl 
& 291 & 02:41:07.19 & +08:44:06.7 & 6563 & 27 \nl 
& 343 & 02:41:53.87 & +08:34:00.0 & 6020 & 27 \nl 
& 344 & 02:41:47.69 & +08:23:04.6 & 6469 & 26 \nl 
& 346 & 02:41:12.81 & +08:43:09.1 & 6044 & 27 \nl 
& 349 & 02:41:08.91 & +08:30:09.9 & 20911 & 59 \nl 
& 351 & 02:41:10.05 & +08:26:31.9 & 6504 & 27 \nl 
& 355 & 02:41:08.29 & +08:12:59.1 & 6492 & 24 \nl 
& 406 & 02:40:38.30 & +08:50:53.7 & 6097 & 34 \nl 
& 422 & 02:40:55.15 & +08:35:44.9 & 5649 & 71 \nl 
& 423 & 02:41:01.08 & +08:31:56.4 & 18531 & 125 \nl 
& 431 & 02:40:17.51 & +08:43:11.0 & 6252 & 39 \nl 
& 440 & 02:39:35.97 & +08:27:51.7 & 12340 & 31 \nl 
MKW2 & & 10:30:14.33 & -03:12:24.2 & \nl 
& 234 & 10:30:17.36 & -03:36:13.2 & 18340 & 99 \nl 
& 236 & 10:30:56.27 & -03:30:34.1 & 11747 & 33 \nl  
& 251 & 10:28:03.42 & -03:40:32.8 & 10791 & 16 \nl
& 252 & 10:27:54.83 & -03:33:30.2 & 9036 & 62 \nl 
& 253 & 10:28:41.82 & -03:37:29.0 & 20925 & 73 \nl 
& 264 & 10:31:23.44 & -03:08:01.1 & 12059 & 23 \nl 
& *275 & 10:31:39.29 & -03:16:00.5 & 28683 & 113 \nl 
& 281 & 10:31:42.92 & -03:24:51.8 & 18444 & 103 \nl 
& 287 & 10:30:28.40 & -03:11:43.5 & 11353 & 29 \nl 
& *288 & 10:30:21.39 & -03:05:52.6 & 10618 & 59 \nl 
& 291 & 10:30:14.60 & -03:12:09.8 & 11346 & 31 \nl 
& 292 & 10:29:54.97 & -03:14:07.5 & 11294 & 119 \nl 
& 295 & 10:30:14.76 & -03:15:40.5 & 11437 & 28 \nl 
& 298 & 10:29:55.82 & -03:20:50.3 & 18489 & 30 \nl 
& 305 & 10:29:33.90 & -03:10:56.4 & 11628 & 56 \nl 
& *306 & 10:30:57.53 & -03:17:35.9 & 6407 & 81 \nl 
& 307 & 10:31:07.16 & -03:14:33.4 & 29313 & 108 \nl 
& *308 & 10:31:00.07 & -03:11:05.5 & 12102 & 122 \nl 
& 309 & 10:30:41.68 & -03:10:59.5 & 10575 & 40 \nl 
& 310 & 10:30:44.68 & -03:10:12.8 & 38152 & 80 \nl 
& 317 & 10:30:26.03 & -03:25:29.0 & 11554 & 45 \nl 
& *318 & 10:31:06.29 & -03:25:24.0 & 11470 & 113 \nl 
& 320 & 10:28:07.14 & -03:16:43.5 & 19902 & 33 \nl 
& 321 & 10:28:06.80 & -03:19:18.2 & 17844 & 39 \nl 
& 323 & 10:28:27.83 & -03:14:34.2 & 10510 & 23 \nl 
& *328 & 10:28:00.97 & -03:20:58.9 & 10474 & 44 \nl 
& 329 & 10:27:49.78 & -03:20:14.6 & 9033 & 50 \nl 
& 339 & 10:27:36.76 & -03:03:59.1 & 15231 & 31 \nl 
& 346 & 10:28:53.66 & -03:15:07.3 & 39602 & 69 \nl 
& 348 & 10:29:00.22 & -03:12:58.1 & 38045 & 101 \nl 
& 351 & 10:29:15.89 & -03:25:37.7 & 13120 & 122 \nl 
& *371 & 10:30:46.91 & -02:45:16.9 & 9006 & 49 \nl 
& 373 & 10:30:28.90 & -02:38:13.4 & 24727 & 122 \nl 
& 379 & 10:30:13.00 & -02:51:37.8 & 29785 & 122 \nl 
& *381 & 10:30:03.43 & -02:49:11.3 & 8809 & 44 \nl 
& 383 & 10:29:23.80 & -02:52:09.6 & 15222 & 35 \nl 
& *387 & 10:30:46.64 & -02:56:18.0 & 15860 & 44 \nl 
& 396 & 10:29:00.84 & -02:57:43.9 & 11187 & 27 \nl  
\tablenotetext{}{$^*$ Emission-Line Redshift}
\enddata
\end{deluxetable}

\clearpage 

\begin{figure}[p]
\plotone{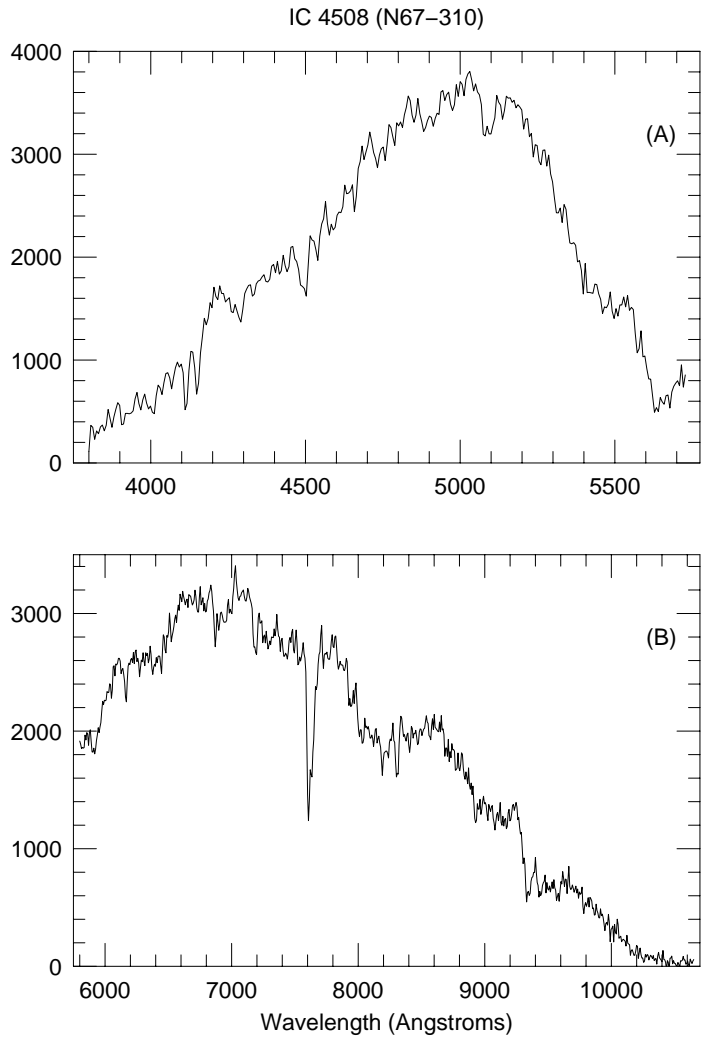}
\caption[]
{
Example spectra from the ARC 3.5-m DIS spectrograph of 
the galaxy IC 4508 in the poor cluster N67-310.  The continuum shape 
is distorted because of the transmission of the dichroic filter 
hear $5750\ang$. a) Blue spectrum, 
b) Red spectrum.
}
\end{figure} 

\clearpage

\begin{figure}[p]
\plotone{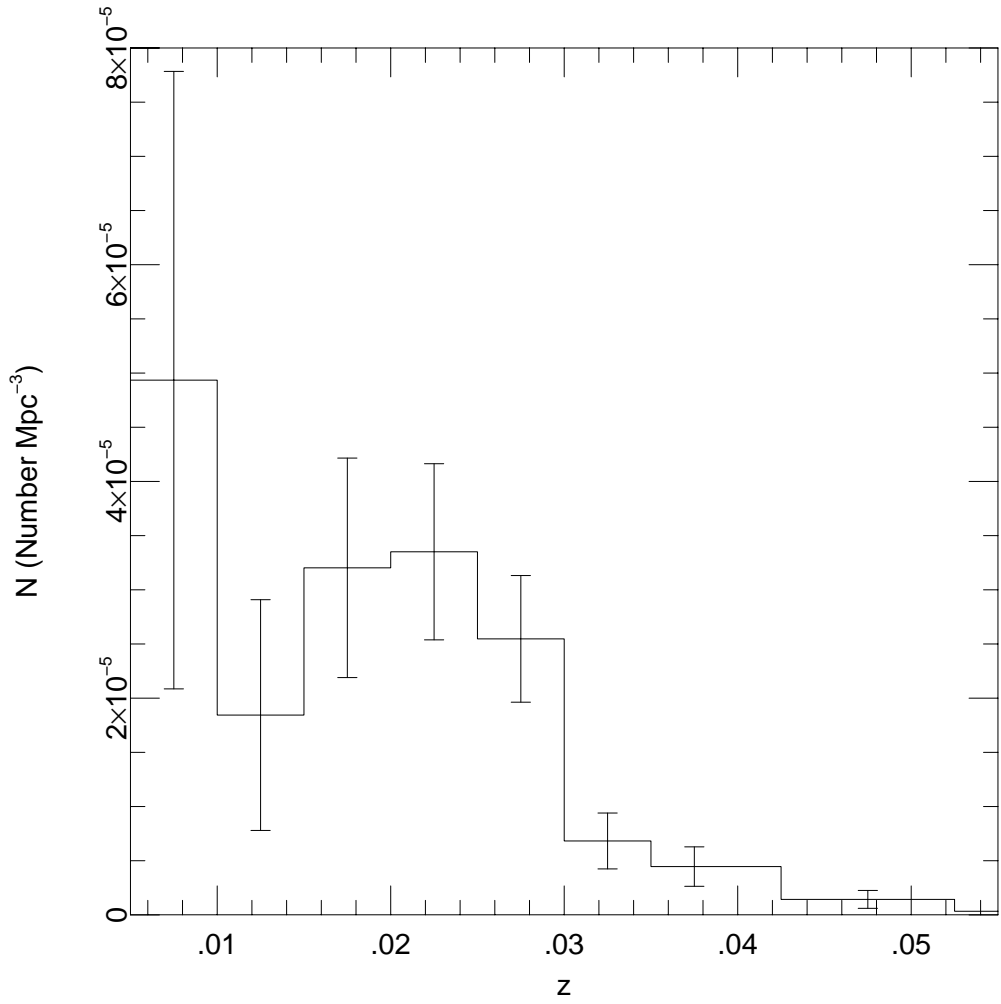} 
\caption
{
The volume-density of our poor cluster sample as a function
of redshift.  We judge the sample nearly complete in the redshift range $0.01\leq z
 \leq 0.03$.
}
\end{figure}

\clearpage

\begin{figure}[p]
\plotone{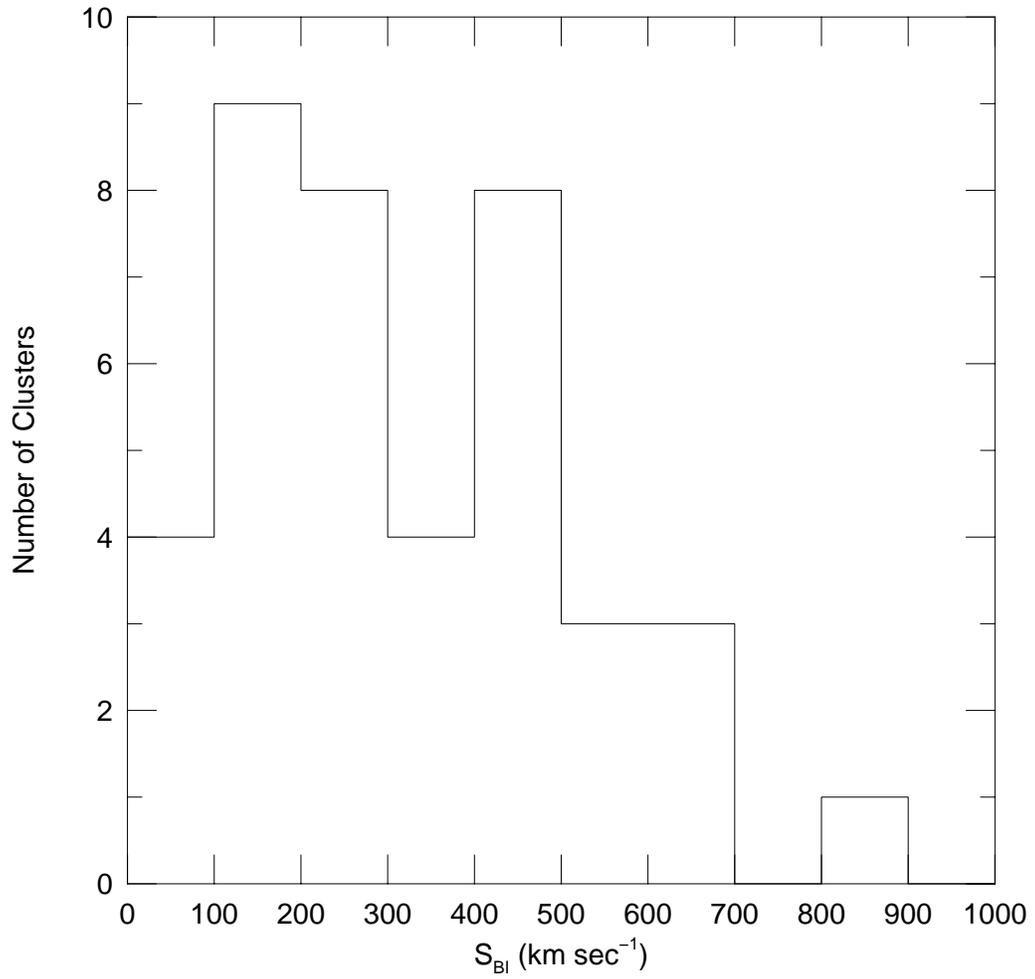}
\caption
{
Histogram of the Biweight scale (velocity dispersion) of all
poor clusters with $\geq 5$ measured velocities within 0.5 Mpc of the
optical cluster center.}
\end{figure}

\clearpage

\begin{figure}[p]
\plotone{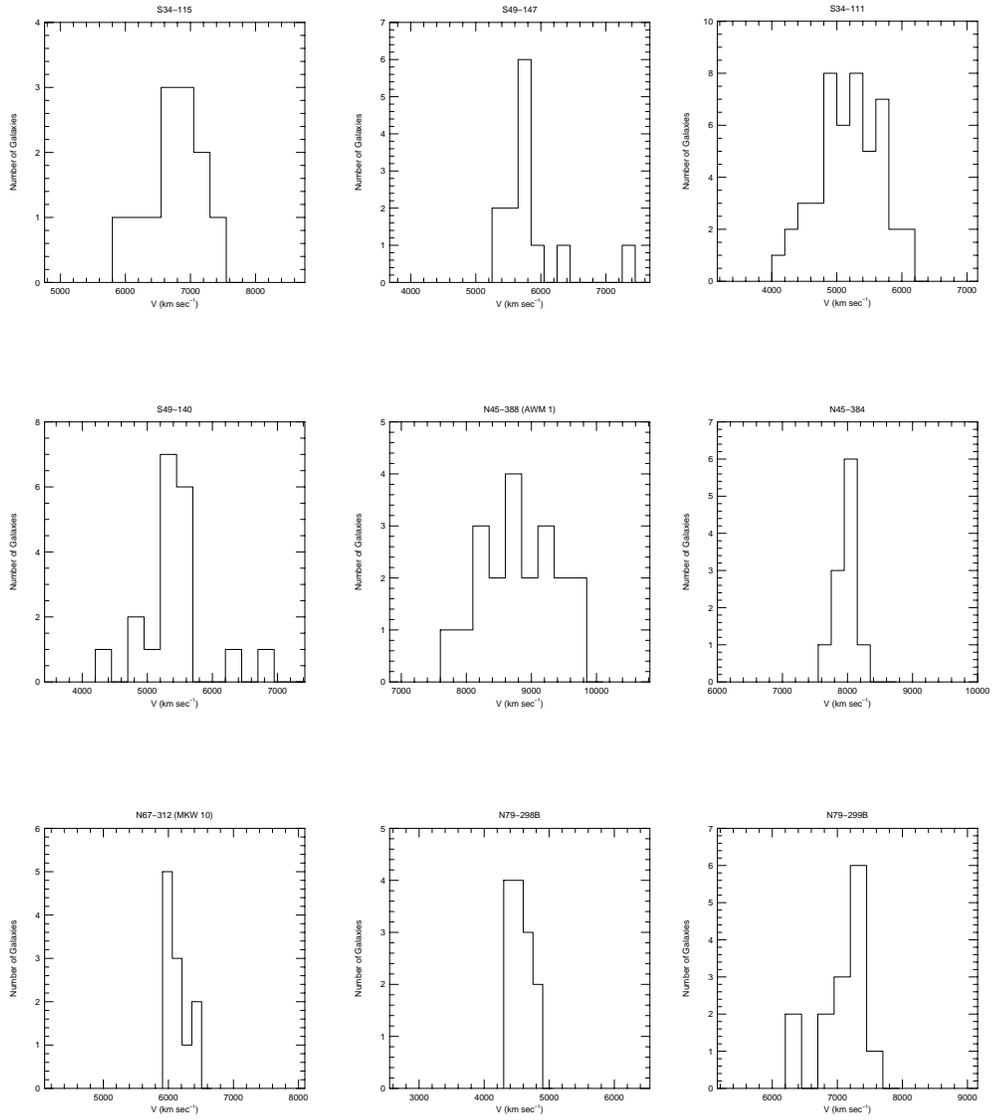}
\caption
{Velocity histograms for all poor clusters with $\geq 10$ 
measured velocities.  All galaxies within $1~Mpc$ and $2000~km~sec^{-1}$
of the cluster mean velocity are plotted.}
\end{figure}

\clearpage

\plotone{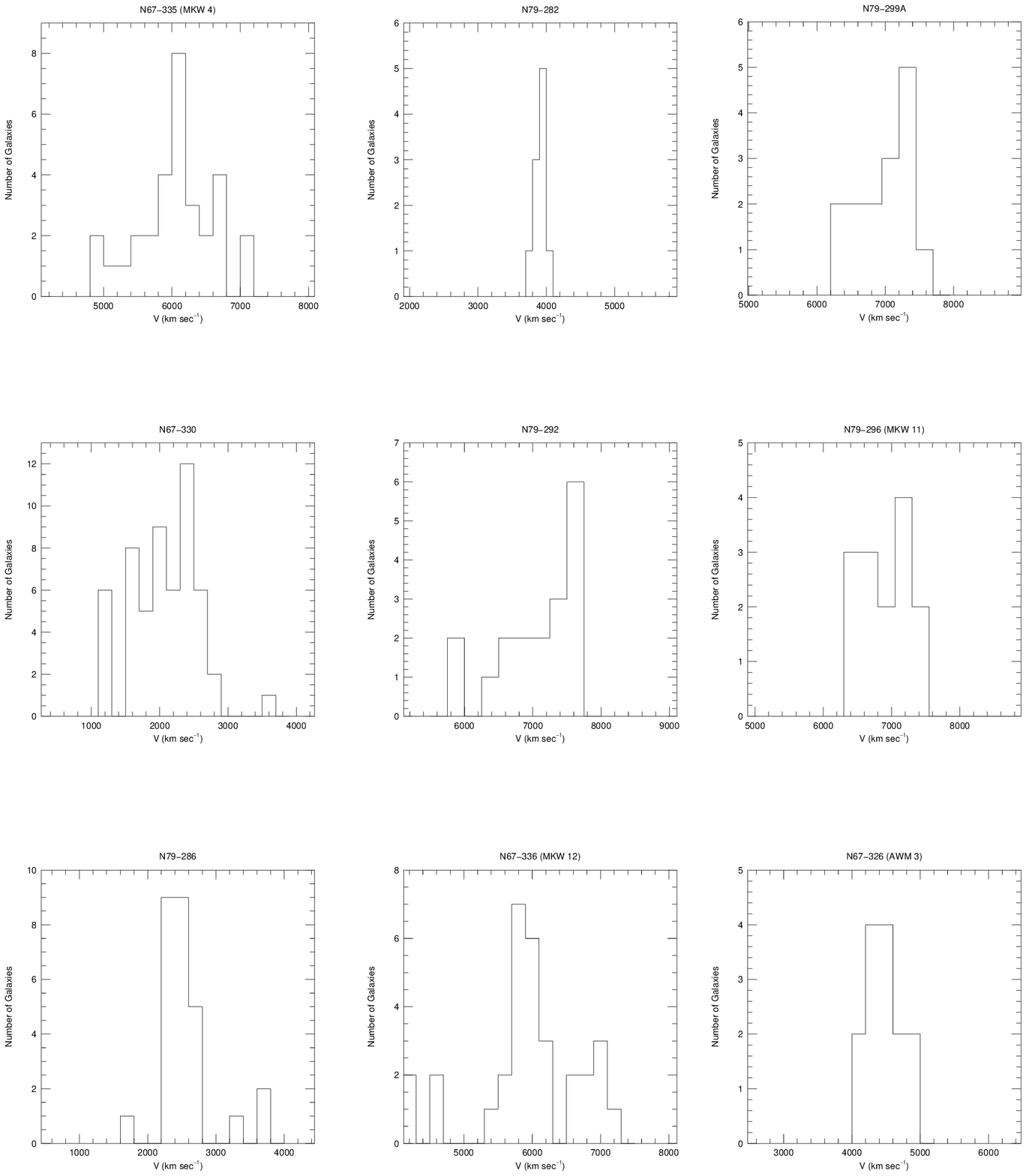}

\clearpage

\plotone{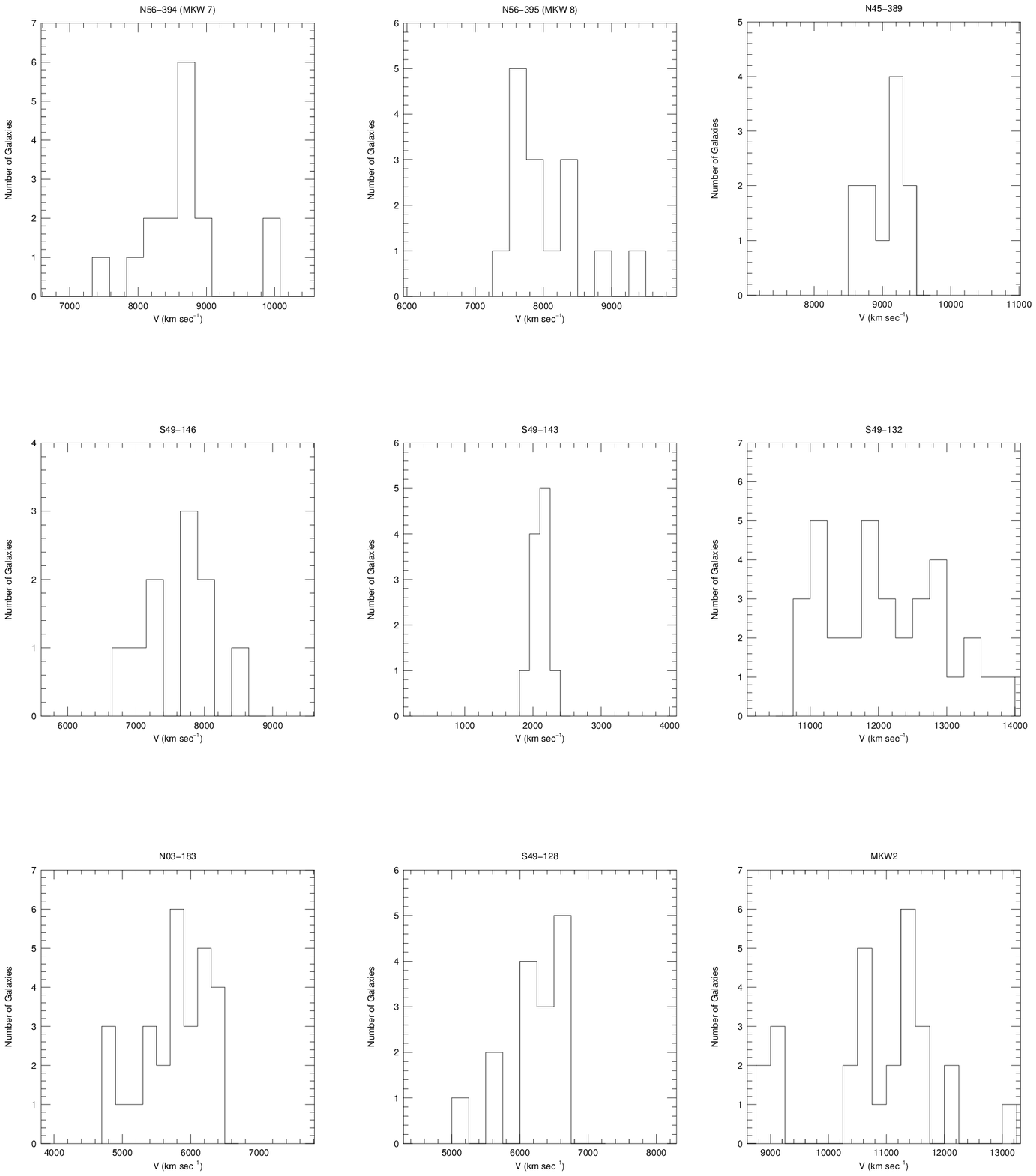}

\clearpage

\begin{figure}[p]
\plotone{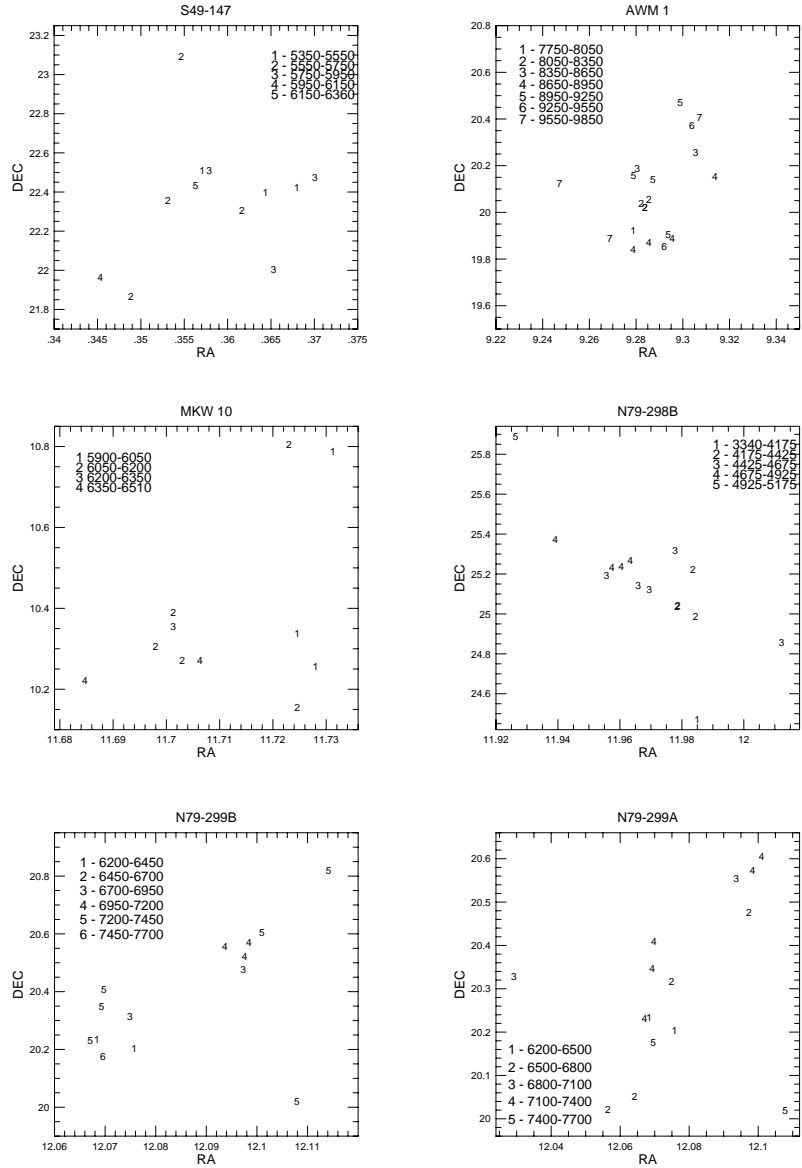}
\caption
{Velocity-coded diagrams of the spatial (RA and DEC) 
galaxy distribution for the 15 clusters with significant non-normality 
as determined from 1-D statistical tests. The velocity coding is labeled
on the diagrams.}   
\end{figure}

\clearpage

\plotone{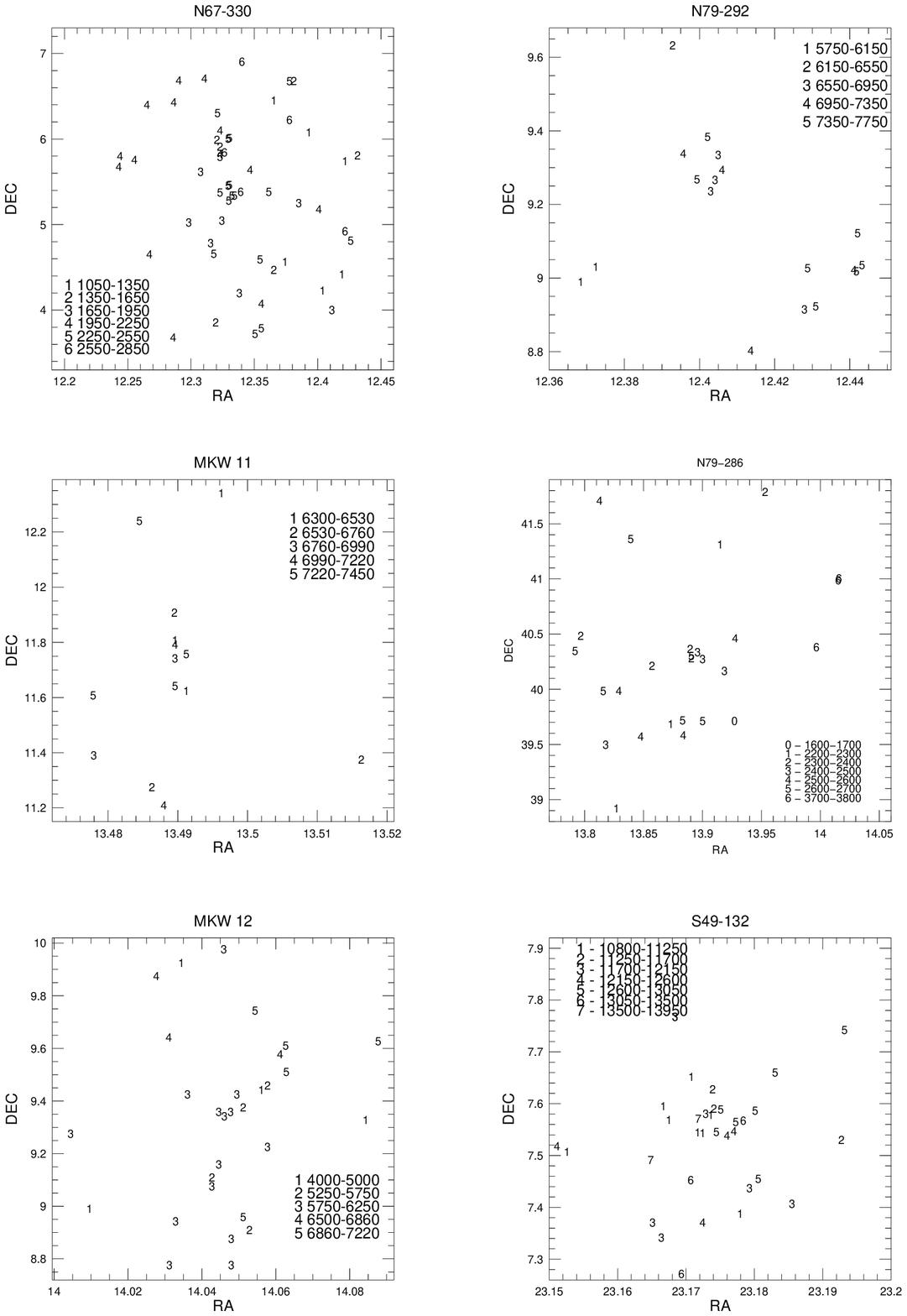}

\clearpage

\plotone{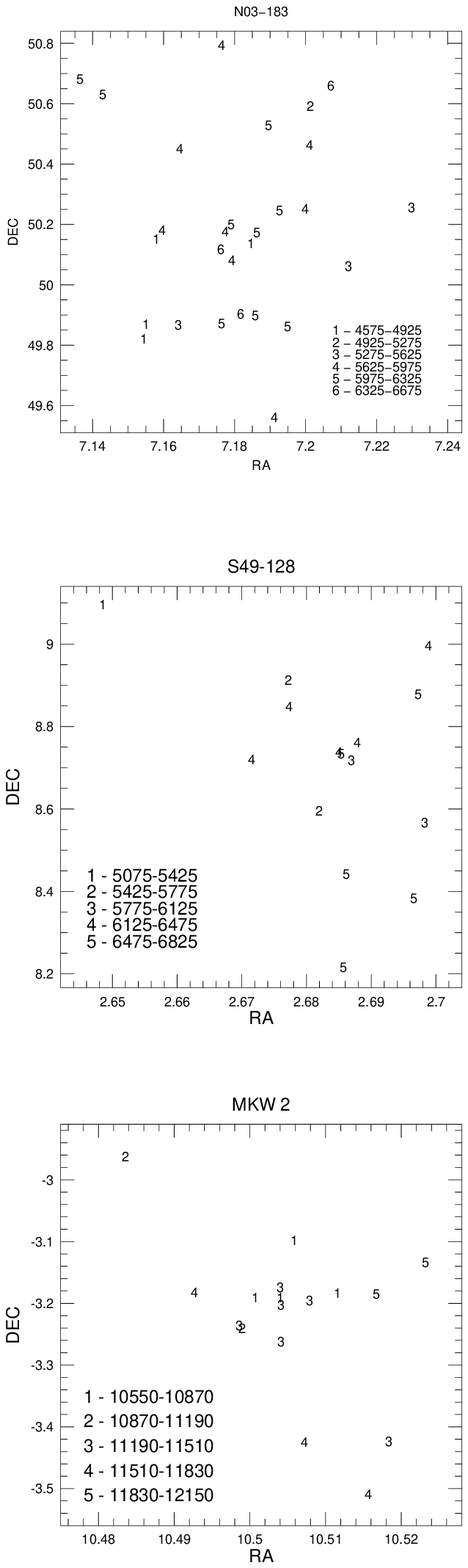}


\begin{references}

\reference{abel58} Abell, G.O. 1958, \apjs, 3, 211

\reference{awm77} Albert, C.E., White, R.A., \& Morgan, W.W. 1977, \apj, 211, 309

\reference{barn85} Barnes, J.E. 1985, \mnras, 215, 517 

\reference{barn86} Barnes, J.E. \& Hut, P. 1986, {\it Nature}, 338, 123 

\reference{beers90} Beers, T.C., Flynn, K., \& Gebhardt, K. 1990, \aj, 100, 32 

\reference{beers95} Beers, T.C., Kriessler, J.R., Bird, C.M., \& Huchra, J.P.
1995, \aj, 109, 874 

\reference{bird93} Bird, C.M. \& Beers, T.C. 1993, \aj, 105, 1596 

\reference{bryan96} Bryan, G. \etal\ 1996, in preparation 

\reference{burn87}  Burns, J.O. Hanisch, R.J., White, R.A., Nelson, E.R., Morrisette, K.A., \& Moody, J.W. 1987, \aj\ 94, 587

\reference{burn96} Burns, J.O., Ledlow, M.J., Loken, C., Klypin, A., Voges, W.,
Bryan, G., \& Norman, M. 1996, in preparation. 

\reference{danese80} Danese, L, DeZotti, G., \& di Tullo, G. 180, \aap, 82, 322 

\reference{david95} David, L.P., Jones, C., \& Forman, W. 1995, \apj, 445, 578 

\reference{dell94} Dell'Antonio, I.P., Geller, M.J., \& Fabricant, D.G. 1994, 
\aj, 107, 427 

\reference{dell95} Dell'Antonio, I.P., Geller, M.J., \& Fabricant, D.G. 1995, 
\aj, 110, 502 

\reference{diaf94} Diaferio, A., Geller, M.J., \& Ramella, M. 1994, \aj, 
107, 868 

\reference{diaf95} Diaferio, A., Geller, M.J., \& Ramella, M. 1995, \aj, 109, 2293 

\reference{doe95} Doe, S.M., Ledlow, M.J., Burns, J.O., \& White, R.A. 1995,
\aj, 110, 46

\reference{dres80} Dressler, A. 1984, \araa, 22, 185 

\reference{dress88} Dressler, A. \& Schectman, S.A. 1988, \aj, 95, 985 

\reference{hern87} Hernquist, L. 1987, \apjs, 64, 715 

\reference{hick82} Hickson, P. 1982, \apj, 255, 382

\reference{hill86} Hill, J.M. \& Lesser, M.P. 1986, in {\it Proceedings 
of SPIE Conference on Instrumentation in Astronomy VI}, {\it SPIE}, 627, 303

\reference{mkw75} Morgan, W.W., Kayser, S., \& White, R.A. 1975, \apj, 199, 545

\reference{most77} Mosteller, F. \& Tukey, J.W. 1977, in {\it Data Analysis and 
Regression} (Addison Wesley, Reading, MA), p. 133

\reference{pink96} Pinkney, J., Roettiger, K., Burns, J.O., \& Bird, C.M. 1996,
\apjs, in press 

\reference{pric91} Price, R., Burns, J.O., Duric, N., \& Newberry, M.V. 1991, \aj, 102, 14 

\reference{Ram94} Ramella, M., Diaferio, A., Geller, M.J., \& Huchra, J.P. 1994, \aj, 107, 1623 

\reference{rose77} Rose, J.A. 1977, \apj, 211, 311 

\reference{shak73} Shakhbazyan, R.K. 1973, {\it Astrofiz}, 9, 495

\reference{son78} Soneira, R.M. \& Peebles, P.J.E. 1978, \apj, 211, 1

\reference{tg76} Turner, E.L. \& Gott, J.R. 1976, \aj, 91, 204

\reference{whit78} White, R.A. 1978, \apj, 226, 591

\reference{wbbb96} White, R.A., Bhavsar, S., Bornmann, P., \& Burns, J.O. 1996, 
in preparation

\reference{yah77} Yahil, A. \& Vidal, N.V. 1977, \apj, 214, 347 

\reference{zab90} Zabludoff, A.I., Huchra, J.P., \& Geller, M.J. 1990, \apjs, 74, 1 

\reference{zwic68} Zwicky, F., Herzog, E., Karpowicz, M., Kowal, C.T., \& Wild, P.
 1961-1968, Catalogue of Galaxies \& Clusters of Galaxies (Caltech, Pasadena)

\end{references}
\end{document}